\documentclass{article}
\usepackage[utf8]{inputenc}
\usepackage[left=3cm,right=3cm,top=3cm,bottom=3cm]{geometry}
\usepackage{amsmath}
\usepackage{amssymb}
\usepackage{graphicx}
\usepackage{esint}
\usepackage{enumitem}
\usepackage{multirow}
\usepackage{graphicx}
\usepackage{hhline}
\usepackage{hyperref}
\usepackage{centernot}
\hypersetup{
  colorlinks   = true,    
  urlcolor     = blue,    
  linkcolor    = blue,    
  citecolor    = black     
}

\newcommand\Tr{\mathrm{Tr}}

\title{One-Loop Effective Action Approach to Quantum MHV Theory}

\author{Hiren Kakkad$^a$,
Piotr Kotko$^a$,
Anna Stasto$^b$
\\ \\
$^a${\it AGH University Of Science and Technology,}\\ 
{\it Faculty of Physics and Applied Computer Science,} \\ 
{\it al. Mickiewicza 30, 30-059 Krakow, Poland} \\ \\
$^b${\it The Pennsylvania State University, Physics Department}\\ 
{\it 104 Davey Lab, University Park, PA 16802, USA }
}

\date{}

\begin{document}
\maketitle

\begin{abstract}

It is well known that the MHV action, i.e. the action containing all the maximally helicity violating vertices, is alone not sufficient for loop computations. In order to develop loop contributions systematically and to  ensure that there are no missing terms, we propose to formulate the quantum MHV action via one-loop effective action approach. The  quadratic field fluctuations in the light cone Yang-Mills theory  are  explicitly integrated,  followed by the classical canonical field  transformation.
We test the approach by calculating one loop $(++++)$ and $(+++-)$ amplitudes, i.e. amplitudes that cannot be calculated from ordinary MHV action. Such an approach can be further used to unambiguously define loop corrections in other theories related to Yang-Mills theory by field transformations.

\end{abstract}

\section{Introduction}
\label{sec:Intro}

Since the discovery that the Maximally Helicity Violating (MHV) tree amplitudes \cite{Parke:1986gb} can be utilized to build other amplitudes \cite{Cachazo2004},
there has been a tremendous progress in our understanding of what might be the most appropriate and effective building blocks of amplitudes. These definitely are not the ordinary Feynman diagrams -- their number grows dramatically, they are not gauge invariant and  -- at loop level -- they often posses singularities that cancel in a sum of diagrams.
Currently, the most effective methods of amplitude calculation are commonly referred to as the "on-shell methods". 
The basic idea is to use on-shell amplitudes themselves as the building blocks. This has a realization either in unitarity methods for loop calculations \cite{Bern1994,Bern2007,Brandhuber2008,Perkins2009,Bern2011}, or as the so-called Britto-Cachazo-Feng-Witten (BCFW) method \cite{Britto:2004ap,Britto:2005fq,Risager2005}, which utilizes on-shell amplitudes with momenta deformed into complex plane as the building blocks. Moreover, it has been recognized, that the on-shell diagrams are part of a yet more general approach to scattering theory, that can be formulated purely in geometrical terms~\cite{Arkani-Hamed_book_2016}.

Despite the fact that the Cachazo-Svrcek-Witten (CSW) method \cite{Cachazo2004} does not give the fewest number of diagrams, its great advantage is that all the building blocks are simply the MHV amplitudes with spinor products continued off-shell; as is known they have an extremely simple form for any number of gluons. It was shown  \cite{Mansfield2006}, that it is possible to formulate a classical action that incorporates the MHV vertices  by applying a canonical field transformation to the light-cone gauge Yang-Mills action (see also \cite{Ettle2006b,Ettle2007,Brandhuber2007a,Brandhuber2007,Ettle2008,Feng2009} for further developments). 
It was also recognized, that the field transformations can be expressed by straight infinite Wilson lines lying on the so-called self-dual plane \cite{Kotko2017,Kakkad2020}.

Due to the presence of only the MHV-type vertices $(+\dots +--)$, amplitudes that have less then two minus helicity are zero in this approach. While at tree level $(+\dots +)$ and $(+\dots +-)$ on-shell amplitudes are indeed zero, this is not the case at loop level \cite{Bern:1991aq,Kunszt_1994}. Indeed, these amplitudes necessarily involve the self-dual vertex $(++-)$ that has been removed by the field redefinition (see eg. \cite{Bardeen1996,Chalmers1996,Cangemi1997,Rosly1997,Monteiro2011} for discussion of the amplitudes in the self-dual sector). 
There have been successful attempts to promote the classical MHV action to the quantum level, focusing however only on the all-plus helicity amplitudes. In particular, in \cite{Ettle2007} the dimensional regularization and careful analysis of the S-matrix equivalence theorem has been used to show the presence of the all-plus amplitudes. Unfortunately, the simple form of the MHV vertices and the field transformation is the feature of the four dimensional theory only. In \cite{Brandhuber2007}, on the other hand, a fully 4D regularization scheme has been used, which requires special counter-terms at one loop. The Authors showed that the self-energy counterterm $(++)$ leads to all-plus one loop  amplitudes in the MHV action, after the field transformation is applied.

In the present work, we suggest another, more general, path to promote the classical MHV theory to the quantum level. The method can be summarized as follows. First, we derive the one loop effective action for the light cone Yang-Mills theory, which therefore accounts for all one loop diagrams of any helicity. Then, we apply the field transformation \cite{Mansfield2006}, to obtain not only the classical MHV action, but also all one loop contributions including the non-MHV one loop diagrams, by construction. 

The significance of our work is that the method can be rather readily generalized to other formulations of gluodynamics, obtained by nonlinear transformation of the classical Yang-Mills fields. 
For example, in \cite{Kakkad:2021uhv} an extension of the MHV action was formulated, that has no triple gluon interactions at all, reducing further the number of diagrams at the tree level as compared to the MHV theory. This new action has also missing loop contributions, similar to the MHV theory.
We expect, that the approach we developed in the present work can be applied to retrieve the missing loop contribution in the new action as well, 
as the one-loop effective action approach seems the universal approach to define such theories at loop level.

Our paper is organized as follows. In Section \ref{sec:SDYM} we review, using the self-dual Yang-Mills theory, the approach of effective action to develop quantum correction. In Section \ref{sec:MHV} we use this approach to develop one-loop quantum corrections to the MHV action, which is the primary result of the present work. In Section \ref{sec:Apps} we use the one-loop effective MHV action to compute $(+ + + +)$ and $(+ + + -)$ one-loop amplitudes validating the method. In Section \ref{sec:conclusion} we summarize the work and discuss future directions. 
 
\section{One-Loop Effective Action for Self-Dual Yang-Mills Theory}
\label{sec:SDYM}

Let us start with a brief review of the approach that will be used to develop quantum corrections to the MHV action. Let us begin with a simple case of the self-dual (SD) Yang-Mills theory. The gauge field in this theory satisfies the self-duality condition
$\hat{F}^{\mu\nu}=\star \hat{F}^{\mu\nu}$, where $\hat{F}^{\mu\nu}=F_a^{\mu\nu}t^a$, with $t^a$ being  generators of the $\mathrm{SU}(N)$ in the fundamental representation\footnote{We use the following normalization of the color generators: $\left[t^{a},t^{b}\right]=i\sqrt{2}f^{abc}t^{c}$ and $\mathrm{Tr}(t^{a}t^{b}) = \delta^{ab}$. To account for the additional factors of $\sqrt{2}$ in this normalization, we re-scale the coupling constant  as $g\rightarrow g/\sqrt{2}$.}. 
It is known that, such a theory can be obtained by a truncation  of the light-cone formulation of the Yang-Mills theory \cite{Chalmers1996}, which contains only ``physical" vertices with an on-shell helicity assignment connected by scalar propagators \cite{Scherk1975} (see also a modern re-derivation in \cite{Schwinn:2005pi}). More precisely, the full Yang-Mills light-cone action contains $(++-)$, $(--+)$ and $(++--)$ helicity vertices (see Section~\ref{subsec:Yang-Mills} for the full action); the SD action is obtained by retaining only the first vertex:
\begin{equation}
    S_{\mathrm{SD}} [A^{\bullet}, A^{\star}] = \int\! dx^{+} d^{3}\mathbf{x}\, \left[
    -\mathrm{Tr}\,\hat{A}^{\bullet}\square\hat{A}^{\star} 
     -2ig\,\mathrm{Tr}\,\partial_{-}^{-1}\partial_{\bullet} \hat{A}^{\bullet}\left[\partial_{-}\hat{A}^{\star},\hat{A}^{\bullet}\right]
    \right]\, ,
    \label{eq:actionSD}
\end{equation}
where we introduced the following double-null coordinates (see \cite{Kotko2017}):
\begin{gather}
v^{+}=v_{\mu}\eta^{\mu}\,, \qquad v^{-}=v_{\mu}\widetilde{\eta}^{\mu}\,, \\
\,\,\,v^{\bullet}=v_{\mu}\varepsilon_{\bot}^{+\,\mu}\,, \qquad v^{\star}=v_{\mu}\varepsilon_{\bot}^{-\,\mu}
\,,\label{eq:double-null_coordinates}
\end{gather}
with basis vectors given by
\begin{gather}
\eta^{\mu}=\frac{1}{\sqrt{2}}\left(1,0,0,-1\right)
\,, \qquad \widetilde{\eta}^{\mu}=\frac{1}{\sqrt{2}}\left(1,0,0,1\right)\,, \\
\varepsilon_{\perp}^{\pm\,\mu}=\frac{1}{\sqrt{2}}\left(0,1,\pm i,0\right)\,.
\label{eq:double-null_basis}
\end{gather}
We also introduced a notation for three-vectors in these coordinates:
\begin{align}
    &\mathbf{x}\equiv\left(x^{-},x^{\bullet},x^{\star}\right)
    \qquad \text{(position space)} \,, \\
    &\mathbf{p}\equiv\left(p^{+},p^{\bullet},p^{\star}\right)
    \qquad \text{(momentum space)} \,.
\end{align}
In these coordinates, the scalar product of two four-vectors reads $u\cdot w=u^{+}w^{-}+u^{-}w^{+}-u^{\bullet}w^{\star}-u^{\star}w^{\bullet}$ and, in particular, $\square=2(\partial_+\partial_- - \partial_{\bullet}\partial_{\star})$.
Notice, that the action \eqref{eq:actionSD} contains just two transverse fields
\begin{gather} 
 {\hat A}^{\bullet}=-\frac{1}{\sqrt{2}} \left({\hat A}^x+i{\hat A}^y\right) \,, \\
 {\hat A}^{\star}=-\frac{1}{\sqrt{2}} \left({\hat A}^x-i{\hat A}^y\right) \,,
\end{gather}
that in the on-shell limit correspond to the two transverse helicity projections. The ${\hat A}^+$ field has been set to zero owing to the light-cone gauge choice, whereas ${\hat A}^-$ is integrated out, or, equivalently, eliminated  using the equation of motion (EOM).

The generating functional for the Green's functions is given as
\begin{equation}
    Z[J]=\int[dA]\, e^{i\left(S_{\mathrm{SD}}[A] + \int\!d^4x\, \Tr \hat{J}_j(x) \hat{A}^j(x)\right) } \,,
\end{equation}
where the index $j$ runs over transverse components $j=\bullet,\star$; the lower index coordinates are defined as $v_{\bullet}=-v^{\star}$ and similar for $v_{\star}$.
In order to derive the one loop-effective action, we shall expand the action around a classical solution ${\hat A}_{c}[J]=({\hat A}_{c}^{\bullet}[J],{\hat A}_{c}^{\star}[J])$ of the EOM: 
\begin{equation}
   \left. \frac{\delta S_{\mathrm{SD}}[A]}{\delta {\hat A}^j(x)}\right|_{{\hat A}={\hat A}_{c}}+{\hat J}_j(x)=0 \,.
    \label{eq:SDEOM0}
\end{equation}
Out of the two equations above, only the equation for ${\hat A}^{\bullet}$ field is dynamical and is often called the self-dual equation:
\begin{equation}
    \Box {\hat{A}}^{\bullet} + 2ig{\partial}_{-} \left[ ({\partial}_{-}^{-1} {\partial}_{\bullet} {\hat{A}}^{\bullet}), {\hat{A}}^{\bullet} \right] + {\hat J}^{\bullet} = 0\, .
    \label{eq:SD_EOM}
\end{equation}
If we take the external currents to be supported on the light cone, the solution ${\hat A}_c^{\bullet}[J]$ provides a generating functional for the Berends-Giele type tree-level off-shell currents \cite{Berends:1987me}, with all gluons having positive helicity, see for example \cite{Cangemi1997}. It also turns out \cite{Kotko2017,Kakkad2020}, that the solution ${\hat A}^{\bullet}_c[J]$ is actually an inverse functional to the 
following, direction-integrated, straight infinite Wilson line
\begin{equation}
-\square^{-1}J_{a}^{\bullet}[A]\left(x\right)=\int_{-\infty}^{\infty}d\alpha\,\mathrm{Tr}\left\{ \frac{1}{2\pi g}t^{a}\partial_{-}\, \mathbb{P}\exp\left[ig\int_{-\infty}^{\infty}ds\, \varepsilon_{\alpha}^+\cdot \hat{A}\left(x+s\varepsilon_{\alpha}^+\right)\right]\right\} \, ,
\label{eq:WilsonLineSol}
\end{equation}
where 
\begin{equation}
    \varepsilon_{\alpha}^{\pm \mu} = \varepsilon_{\perp}^{\pm \mu}- \alpha \eta^{\mu} \, .
    \label{eq:epsilon_alpha}
\end{equation}
This fact is related to the integrability of the self-dual theory.

Up to the second order in fields, the expansion of the action around the classical solution ${\hat A}_c[J]$ reads
\begin{multline}
    S_{\mathrm{SD}}[A] + \int\!d^4x\, \Tr \hat{J}_i(x) \hat{A}^i(x)  
    = S_{\mathrm{SD}}[A_c] + \int\!d^4x\, \Tr \hat{J}_i(x)\hat{A}_c^i(x) \\ + \int\!d^4x\,\Tr\left(\hat{A}^i(x)-\hat{A}_c^i(x)\right)
    \left(\frac{\delta S_{\mathrm{SD}}[A_c]}{\delta \hat{A}^i(x)}+{\hat J}_i(x)\right) \\
    +\frac{1}{2}\int\!d^4xd^4y\,\Tr\left(\hat{A}^i(x)-\hat{A}_c^i(x)\right)\frac{\delta^2 S_{\mathrm{SD}}[A_c]}{\delta\hat{A}^i(x)\delta\hat{A}^j(y)}\left(\hat{A}^j(y)-\hat{A}_c^j(y)\right) + \dots
\end{multline}
The linear term is zero due to the EOM \eqref{eq:SDEOM0}, while the integration over the quadratic term in the partition function gives
\begin{equation}
    Z_{\mathrm{SD}}[J]\approx \left[\det
    \left( 
    \frac{\delta^2 S_{\mathrm{SD}}[A_c]}
    {\delta \hat{A}^i(x)\delta \hat{A}^j(y)}
    \right)\right]^{-\frac{1}{2}}
    \exp\left\{i\left(S_{\mathrm{SD}}[A_c] 
    + \int\!d^4x\, \Tr \hat{J}_i(x) \hat{A}_c^i(x) \right) \right\} \,.
\end{equation}
The functional determinant can be slightly simplified owing to the fact that the action is linear in ${\hat A}^{\star}$:
\begin{equation}
    \sqrt{\det
    \left( 
    \frac{\delta^2 S_{\mathrm{SD}}[A_c]}
    {\delta \hat{A}^i(x)\delta \hat{A}^j(y)}
    \right)}
    =  \det
    \left( 
    \frac{\delta^2 S_{\mathrm{SD}}[A_c]}
    {\delta \hat{A}^{\star}(x)\delta \hat{A}^{\bullet}(y)}
    \right)
     \,.
\end{equation}
Using the known representation for a functional determinant in terms of an exponential of a logarithm, we get
\begin{equation}
    Z_{\mathrm{SD}}[J]\approx 
    \exp\left\{ iS_{\mathrm{SD}}[A_c] 
    + i\int\!d^4x\, \Tr \hat{J}_i(x) \hat{A}_c^i(x) 
    - \Tr\ln\left(\frac{\delta^2 S_{\mathrm{SD}}[A_c]}
    {\delta \hat{A}^{\star}(x)\delta \hat{A}^{\bullet}(y)} \right) 
    \right\}
    \label{eq:Partition1}
    \,,
\end{equation}
where the second trace goes over all degrees of freedom of the matrix under the logarithm, including the positions.
In order to simplify the notation, in what follows, we shall use a very convenient "collective index" convention. That is, we shall use a single index $I,J,K,\dots$ to collectively denote color indices, position, etc. Summation over such an index thus corresponds to both summation and integration. Using that notation, the argument of the logarithm reads
\begin{equation}
   \frac{\delta^2 S_{\mathrm{SD}}[A_c]}
    {\delta A^{\star I}\delta A^{\bullet J}} 
    = -\square_{IJ}-\left(V_{-++}\right)_{IJK}A_{c}^{\bullet K}
    \label{eq:d2SdAdA}
    \,.
\end{equation}
 The vertex above is a shorthand notation for the $(-++)$ helicity triple gluon vertex, which in our normalization reads 
\begin{equation}
    V_{-++}^{abc}(x,y,z)=2gf^{abc}\delta^4(x-y)
    \delta^4(x-z)\left(\omega_y-\omega_z\right)\partial_{-}(x) \, ,
    \label{eq:v3_position}
\end{equation}
where the  differential operator $\omega_x$ is defined as
\begin{equation}
    \omega_x=\partial^{-1}_{-}(x)\partial_{\bullet}(x) \,.
\end{equation}
In the brackets on the r.h.s we denote variables on which the derivatives act.

\begin{figure}
    \centering
    \includegraphics[width=9cm]{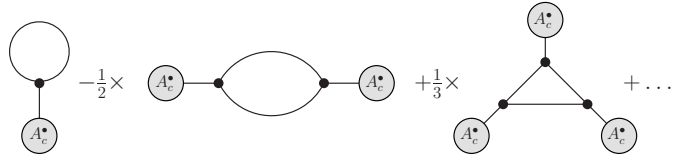}
    \caption{
    \small
    The diagrammatic content of the logarithm term in the one-loop effective action for the self-dual Yang-Mills theory.}
    \label{fig:SD_oneloopEffaction}
\end{figure}

Equation \eqref{eq:Partition1} provides an approximation to the full generating functional, taking into account all one-loop contributions.
Let us now briefly recall, how one makes this fact explicit.
First, we factor out the inverse propagator in \eqref{eq:d2SdAdA}, which together with the logarithm constitutes an (inverse) determinant $\det(\square)$, which we disregard from the generating functional. Expanding the remaining logarithm and focusing on the generating functional for the connected contributions, we have
\begin{equation}
    W_{\mathrm{SD}}[J]=S_{\mathrm{SD}}[A_c]+J_I A_c^{\bullet I}
    +i\,\Tr \sum_{k=1}^{\infty} \frac{(-1)^{k+1}}{k} 
    \left[\square^{-1}_{IJ}\left(V_{-++}\right)_{JKL}A_{c}^{\bullet L}\right]^k \,.
    \label{eq:W_functional}
\end{equation}
One should remember, that the ${\hat A}^{\bullet}$ fields above are the classical solutions depending on the external currents ${\hat J}$, ${\hat A}_c={\hat A}_c[J]$. Since we are interested in one-loop contributions, we make the Legendre transform
\begin{equation}
    \Gamma_{\mathrm{SD}}[A_c]=W_{\mathrm{SD}}[J]-J_IA_c^{\bullet I} \,,
\end{equation}
which renders the classical fields independent of external "currents" and introduces a new generating functional $\Gamma[A_c]$ called one-loop effective action, because it generates all contributions up to one-loop:  \begin{align}
    \Gamma_{\mathrm{SD}}[A_c] =& S_{\mathrm{SD}}[A_c]
    +i \, \Tr \sum_{k=1}^{\infty} \frac{(-1)^{k+1}}{k} 
    \left[\square^{-1}_{IJ} \left(V_{-++}\right)_{JKL} A_c^{\bullet L}\right]^k \nonumber\\
    =&- A_c^{\star I}\square_{IJ}A_c^{\bullet J} 
    - \frac{1}{2}\left(V_{-++}\right)_{IJK}A_c^{\star I}A_c^{\bullet J}A_c^{\bullet K} \nonumber\\
    &+i\, \square^{-1}_{IJ}\left(V_{-++}\right)_{JIK} A_c^{\bullet K}
    -i\, \frac{1}{2} \square^{-1}_{I_1J_1}\left(V_{-++}\right)_{J_1I_2K_1}
    \square^{-1}_{I_2J_2}\left(V_{-++}\right)_{J_2I_1K_2} 
    A_c^{\bullet K_1} A_c^{\bullet K_2} + \dots \,.
\end{align}
In the last line we explicitly show the tadpole and self energy-type diagram (see Fig.~\ref{fig:SD_oneloopEffaction}).

An immediate application of the one-loop effective action derived above is a direct and simple proof of the method \cite{Brandhuber2007a} for calculating one-loop all-plus-helicity amplitudes.
In that work, the Authors considered a \emph{holomorphic} field transformation
\begin{gather}
    {\hat A}^{\bullet} \longrightarrow {\hat B}^{\bullet}={\hat B}^{\bullet}[A^{\bullet}] \,, \nonumber  \\
    {\hat A}^{\star} \longrightarrow {\hat B}^{\star}={\hat A}^{\star} \,,
    \label{eq:BrandhubeTransf}
\end{gather}
where the first transformation had exactly the same form as in \cite{Mansfield2006}. In the latter work, which we review in more detail later in Section~\ref{subsec:MHV},  field redefinitions have been applied to the full Yang-Mills action to obtain the classical MHV action. There, the second transformation -- for the $\star$ field -- was chosen in such a way that the transformation is \emph{canonical} and thus, effectively, there is no Jacobian for the field transformation.
In \cite{Kotko2017} it has been shown, that the transformation for the $\bullet$ field is actually given by the same Wilson line as in \eqref{eq:WilsonLineSol}:
\begin{equation}
    B^{\bullet}_{a}[A^{\bullet}]\left(x\right)=\int_{-\infty}^{\infty}d\alpha\,\mathrm{Tr}\left\{ \frac{1}{2\pi g}t^{a}\partial_{-}\, \mathbb{P}\exp\left[ig\int_{-\infty}^{\infty}ds\,  \hat{A}^{\bullet}\left(x+s\varepsilon_{\alpha}^+\right)\right]\right\} \, ,
\label{eq:WilsonLineBbullet}
\end{equation}
where we used the fact that $\varepsilon_{\alpha}^+\cdot \hat{A}={\hat A}^{\bullet}$ in light-cone gauge.
Applying the transformation \eqref{eq:BrandhubeTransf} to the generating functional we have
\begin{equation}
    Z_{\mathrm{SD}}[J] = \int[dB] \det\left(\frac{\delta A^{\bullet I}}{\delta B^{\bullet J}}\right) \exp\left\{ iS_{\mathrm{SD}}[B^{\bullet},B^{\star}] 
    + i\,A^{\bullet I}[B^{\bullet}]J_{\bullet I}
    + i\,B^{\star K}J_{\star K}\right\} \,,
    \label{eq:BrandhuberZ}
\end{equation}
where ${\hat A}^{\bullet}[B^{\bullet}]$ is an inverse functional to \eqref{eq:WilsonLineBbullet}, which was first calculated in \cite{Ettle2006b} and next obtained in the Wilson-line-like form in \cite{Kakkad2020}. Notice, that
\begin{equation}
    {\hat A}^{\bullet}[B^{\bullet}]={\hat A}^{\bullet}_c[J] \,.
    \label{eq:InverseWL=Ac}
\end{equation}
with ${\hat J}^{\bullet}= - \square {\hat B}^{\bullet}$. As demonstrated in \cite{Brandhuber2007a}, the Jacobian in \eqref{eq:BrandhuberZ} gives rise to the all-plus helicity one-loop amplitudes.
In order to see this from the one-loop effective action, let us go back to Eq.~\eqref{eq:Partition1} and note that, from \eqref{eq:SDEOM0} we have 
\begin{equation}
    \frac{\delta^2 S[A_c]}{\delta {\hat A}^{\bullet}(y)\delta {\hat A}^{\star}(x)}=\frac{\delta {\hat J}^{\bullet}[A_c^{\bullet}](x)}{\delta {\hat A}^{\bullet}(y)} \,,
\end{equation}
where we used ${\hat J}_{\star}=-{\hat J}^{\bullet}$. Thus, to one-loop, 
\begin{multline}
    Z_{\mathrm{SD}}[J]\approx \left[\det\left( \frac{\delta {\hat J}^{\bullet}[A_c^{\bullet}](x)}{\delta {\hat A}^{\bullet}(y)} \right)\right]^{-1} 
     \exp\left\{ iS_{\mathrm{SD}}[A_c] 
    + i\int\!d^4x\, \Tr \hat{J}_i(x) \hat{A}_c^i(x) \right\}  \\
    = \det\left( \frac{\delta {\hat A}_c^{\bullet}[J](y)}{\delta {\hat B}^{\bullet}(x)} \right)
     \exp\left\{ iS_{\mathrm{SD}}[A_c] 
    + i\int\!d^4x\, \Tr \hat{J}_i(x) \hat{A}_c^i(x) \right\} \,.
\end{multline}
Above, we substituted ${\hat J}^{\bullet}= - \square {\hat B}^{\bullet}$ in the first line and discarded the $\det(\square)$ factor to obtain the second line.
We see, that  the equation following from the holomorphic field redefinition \eqref{eq:BrandhuberZ} and the one above, which follows from the one-loop effective action, generate exactly the same one loop amplitudes.

\section{One-Loop Effective Action for MHV vertices}
\label{sec:MHV}

In order to obtain the MHV action which includes all the 
ingredients necessary to obtain one-loop amplitudes, we shall 
apply the following strategy. First, we derive the one-loop
effective action for the full Yang-Mills theory on the light cone.
Next, we apply the field transformations to obtain the MHV action and all the required one loop diagrams. Interestingly, as we shall see, this procedure does not require evading the S-matrix equivalence theorem as in \cite{Ettle2007} or using special counter terms to generate the missing contributions as in \cite{Brandhuber2007}. 

\subsection{One-loop Effective action for Yang-Mills theory on the light cone}
\label{subsec:Yang-Mills}

The full Yang-Mill action on the light cone includes, in addition to the self-dual part, also $(--+)$ and $(--++)$ helicity vertices. Using the notation from the previous section it reads
\begin{multline}
S_{\mathrm{YM}}\left[A^{\bullet},A^{\star}\right]=\int dx^{+}\int d^{3}\mathbf{x}\,\,\Bigg\{ 
-\mathrm{Tr}\,\hat{A}^{\bullet}\square\hat{A}^{\star}
-2ig\,\mathrm{Tr}\,\partial_{-}^{-1}\partial_{\bullet} \hat{A}^{\bullet}\left[\partial_{-}\hat{A}^{\star},\hat{A}^{\bullet}\right] \\
-2ig\,\mathrm{Tr}\,\partial_{-}^{-1}\partial_{\star}\hat{A}^{\star}\left[\partial_{-}\hat{A}^{\bullet},\hat{A}^{\star}\right]
-2g^{2}\,\mathrm{Tr}\,\left[\partial_{-}\hat{A}^{\bullet},\hat{A}^{\star}\right]\partial_{-}^{-2}\left[\partial_{-}\hat{A}^{\star},\hat{A}^{\bullet}\right]
\Bigg\}
\,.\label{eq:YM_LC_action}
\end{multline}
Applying similar steps as we did for the self-dual theory (i.e. expanding around a classical solution and integrating the field fluctuations) we get
\begin{equation}
    Z_{\mathrm{YM}}[J]\approx \left[\det
    \left( 
    \frac{\delta^2 S_{\mathrm{YM}}[A_c]}
    {\delta \hat{A}^i(x)\delta \hat{A}^j(y)}
    \right)\right]^{-\frac{1}{2}}
    \exp\left\{i\left(S_{\mathrm{YM}}[A_c] 
    + \int\!d^4x\, \Tr \hat{J}_i(x) \hat{A}_c^i(x) \right) \right\} \,,
    \label{eq:Par_det}
\end{equation}
where, using the collective indices introduced above, the functional determinant reads:
\begin{equation}
    \det  \left( 
    \frac{\delta^2 S_{\mathrm{YM}}[A_c]}
    {\delta \hat{A}^i(x)\delta \hat{A}^j(y)}
    \right)  = \det \begin{vmatrix}
     \frac{\delta^2 S_{\mathrm{YM}}[A_c]}
    {\delta A^{\star I}\delta A^{\bullet J}} 
     & \frac{\delta^2 S_{\mathrm{YM}}[A_c]}
    {\delta A^{\bullet I}\delta A^{\bullet J}} \\ \\
\frac{\delta^2 S_{\mathrm{YM}}[A_c]}
    {\delta A^{\star I}\delta A^{\star J}} & \frac{\delta^2 S_{\mathrm{YM}}[A_c]}
    {\delta A^{\bullet I}\delta A^{\star J}} 
    \label{eq:DET}
\end{vmatrix} \,.
\end{equation}
Note, the operators in the block matrix above are  non-commuting. Therefore the usual definition for the determinant of $2\times2$ matrix does not hold. However, since the block sizes are the same, the determinant can alternatively be expressed as 
\begin{equation}
   \det \left[ \frac{\delta^2 S_{\mathrm{YM}}[A_c]}
    {\delta A^{\star I}\delta A^{\bullet K}} \, \frac{\delta^2 S_{\mathrm{YM}}[A_c]}
    {\delta A^{\star K}\delta A^{\bullet J}} - \frac{\delta^2 S_{\mathrm{YM}}[A_c]}
    {\delta A^{\star I}\delta A^{\bullet K}} \, \frac{\delta^2 S_{\mathrm{YM}}[A_c]}
    {\delta A^{\star K}\delta A^{\star L}} \left( \frac{\delta^2 S_{\mathrm{YM}}[A_c]}
    {\delta A^{\bullet L}\delta A^{\star M}} \right)^{-1} \frac{\delta^2 S_{\mathrm{YM}}[A_c]}
    {\delta A^{\bullet M}\delta A^{\bullet J}}\right]\,.
    \label{eq:determinant}
\end{equation}
Above, the same indices are contracted. Since the collective indices consist of both the discrete and continuous parameters, contracted indices result in a sum for the discrete quantities and an integral for the continuous ones. The matrices in Eq.~\eqref{eq:determinant} read:
\begin{equation}
   \frac{\delta^2 S_{\mathrm{YM}}[A_c]}
    {\delta A^{\star I}\delta A^{\bullet J}} 
    = -\square_{IJ}-\left(V_{-++}\right)_{IJK}A_{c}^{\bullet K} -\left(V_{--+}\right)_{KIJ}A_{c}^{\star K} -\left(V_{--++}\right)_{LIJK}A_{c}^{\star L} A_{c}^{\bullet K}
    \label{eq:S+-}
    \,,
\end{equation}
\begin{equation}
   \frac{\delta^2 S_{\mathrm{YM}}[A_c]}
    {\delta A^{\bullet I}\delta A^{\bullet J}} 
    = -\left(V_{-++}\right)_{KIJ}A_{c}^{\star K} -\left(V_{--++}\right)_{KLIJ}A_{c}^{\star K} A_{c}^{\star L}
    \label{eq:S++}
    \,,
\end{equation}
\begin{equation}
   \frac{\delta^2 S_{\mathrm{YM}}[A_c]}
    {\delta A^{\star I}\delta A^{\star J}} 
    = -\left(V_{--+}\right)_{IJK}A_{c}^{\bullet K} -\left(V_{--++}\right)_{IJKL}A_{c}^{\bullet K} A_{c}^{\bullet L}
    \label{eq:S--}
    \,,
\end{equation}

Substituting \eqref{eq:determinant} in the partition function and replacing the determinant with an exponential of a trace of a logarithm, we get:
\begin{multline}
    Z_{\mathrm{YM}}[J]\approx 
    \exp\left\{ iS_{\mathrm{YM}}[A_c] 
    + i\int\!d^4x\, \Tr \hat{J}_i(x) \hat{A}_c^i(x) \right.\\
   \left. - \frac{1}{2} \Tr\ln \left[ \frac{\delta^2 S_{\mathrm{YM}}[A_c]}
    {\delta A^{\star I}\delta A^{\bullet K}} \, \frac{\delta^2 S_{\mathrm{YM}}[A_c]}
    {\delta A^{\star K}\delta A^{\bullet J}} - \frac{\delta^2 S_{\mathrm{YM}}[A_c]}
    {\delta A^{\star I}\delta A^{\bullet K}} \, \frac{\delta^2 S_{\mathrm{YM}}[A_c]}
    {\delta A^{\star K}\delta A^{\star L}} \left( \frac{\delta^2 S_{\mathrm{YM}}[A_c]}
    {\delta A^{\bullet L}\delta A^{\star M}} \right)^{-1} \frac{\delta^2 S_{\mathrm{YM}}[A_c]}
    {\delta A^{\bullet M}\delta A^{\bullet J}}\right]
    \right\}
    \label{eq:Partition2}
    \,.
\end{multline}

\begin{figure}
    \centering
    \includegraphics[width=15cm]{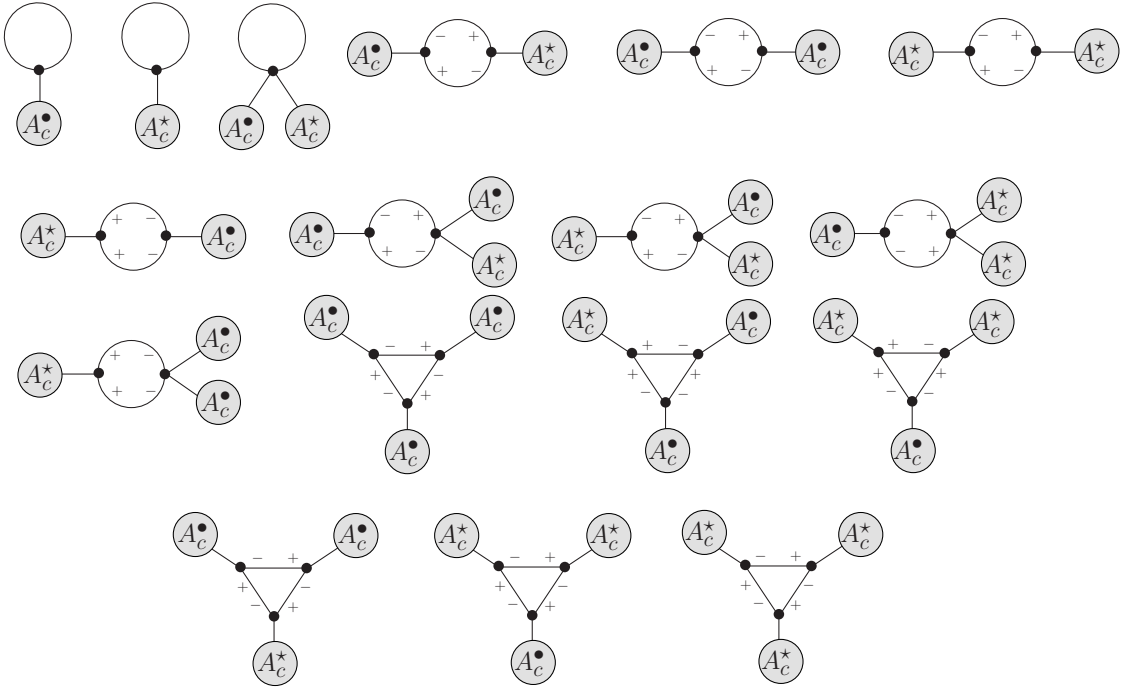}
    \caption{
    \small
    The diagrammatic content, up to 3-point, of the logarithm term in one-loop effective action for the Yang-Mills theory. For simplicity, we suppressed the symmetry factors and signs associated with each diagram.
    The external lines with blobs represent ${\hat A}^{\bullet}_c$ or ${\hat A}^{\star}_c$ field attachments.
    }
    \label{fig:YM_oneloopEffaction}
\end{figure}

The one-loop term here is much more complicated then in the self-dual case. Let us start with factoring out the $\det(\square)$. At first sight it may appear that we get $(\square)^2$ as evident from the first term in the log. But the factor of one-half outside the log takes care of that. With a bit of algebra one can see that a similar factor arises from the second term in the log. 

After factoring out the inverse propagator, the explicit expansion of the logarithm term up to 2-point reads:
\begin{align}
   \frac{i}{2}\, \Bigg[2\, \square^{-1}_{IJ}\left(V_{-++}\right)_{JIK} A_c^{\bullet K} &+2\, \square^{-1}_{IJ}\left(V_{-+-}\right)_{JIK} A_c^{\star K} +2\, \square^{-1}_{IJ}\left(V_{-++-}\right)_{JIKL} A_c^{\bullet K}A_c^{\star L} \nonumber\\
    &+\square^{-1}_{I_1J_1}\left(V_{-++}\right)_{J_1I_2K_1}
    \square^{-1}_{I_2J_2}\left(V_{-++}\right)_{J_2I_1K_2} 
    A_c^{\bullet K_1} A_c^{\bullet K_2}\nonumber\\
    &+\square^{-1}_{I_1J_1}\left(V_{-+-}\right)_{J_1I_2K_1}
    \square^{-1}_{I_2J_2}\left(V_{-+-}\right)_{J_2I_1K_2} 
    A_c^{\star K_1} A_c^{\star K_2}\nonumber\\
    &+\,\square^{-1}_{I_1J_1}\left(V_{-++}\right)_{J_1I_2K_1}
    \square^{-1}_{I_2J_2}\left(V_{-+-}\right)_{J_2I_1K_2} 
    A_c^{\bullet K_1} A_c^{\star K_2}\nonumber\\
    &+\,\square^{-1}_{I_1J_1}\left(V_{-+-}\right)_{J_1I_2K_1}
    \square^{-1}_{I_2J_2}\left(V_{-++}\right)_{J_2I_1K_2} 
    A_c^{\star K_1} A_c^{\bullet K_2}\nonumber\\
    &-\,\square^{-1}_{I_1J_1}\left(V_{++-}\right)_{J_1I_2K_1}
    \square^{-1}_{I_2J_2}\left(V_{--+}\right)_{J_2I_1K_2} 
    A_c^{\star K_1} A_c^{\bullet K_2}+ \dots \,\nonumber\\
    &-\,\frac{1}{2} \Big\{ 4\,\square^{-1}_{I_1J_1}\left(V_{-++}\right)_{J_1I_2K_1}
    \square^{-1}_{I_2J_2}\left(V_{-++}\right)_{J_2I_1K_2} 
    A_c^{\bullet K_1} A_c^{\bullet K_2} \nonumber\\
    &+4\,\square^{-1}_{I_1J_1}\left(V_{-+-}\right)_{J_1I_2K_1}
    \square^{-1}_{I_2J_2}\left(V_{-+-}\right)_{J_2I_1K_2} 
    A_c^{\star K_1} A_c^{\star K_2}\nonumber\\
    &+4\,\,\square^{-1}_{I_1J_1}\left(V_{-++}\right)_{J_1I_2K_1}
    \square^{-1}_{I_2J_2}\left(V_{-+-}\right)_{J_2I_1K_2} 
    A_c^{\bullet K_1} A_c^{\star K_2}\nonumber\\
    &+4\,\,\square^{-1}_{I_1J_1}\left(V_{-+-}\right)_{J_1I_2K_1}
    \square^{-1}_{I_2J_2}\left(V_{-++}\right)_{J_2I_1K_2} 
    A_c^{\star K_1} A_c^{\bullet K_2} + \dots \, \Big\}
    \Bigg]\,.
    \label{eq:2p_1loopexp}
\end{align}
The terms above include tadpoles and bubbles with varying helicities.  The expansion produces appropriate combinatorial factors necessary to compute the amplitudes. This is considered in detail in Section~\ref{sec:Apps}, where we compute amplitudes. In Figure~\ref{fig:YM_oneloopEffaction}, the expansion up to 3-point is displayed omitting the symmetry factors and signs associated with each contribution for brevity.  Notice, each contraction happens over opposite helicities due to the kinetic term of the action.

The one-loop effective action, via the Legendre transform of the generating functional for the connected Green's function, reads
\begin{multline}
    \Gamma_{\mathrm{YM}}[A_c] = S_{\mathrm{YM}}[A_c] \\
    + \frac{i}{2} \Tr\ln \left[ \frac{\delta^2 S_{\mathrm{YM}}[A_c]}
    {\delta A^{\star I}\delta A^{\bullet K}} \, \frac{\delta^2 S_{\mathrm{YM}}[A_c]}
    {\delta A^{\star K}\delta A^{\bullet J}} - \frac{\delta^2 S_{\mathrm{YM}}[A_c]}
    {\delta A^{\star I}\delta A^{\bullet K}} \, \frac{\delta^2 S_{\mathrm{YM}}[A_c]}
    {\delta A^{\star K}\delta A^{\star L}} \left( \frac{\delta^2 S_{\mathrm{YM}}[A_c]}
    {\delta A^{\bullet L}\delta A^{\star M}} \right)^{-1} \frac{\delta^2 S_{\mathrm{YM}}[A_c]}
    {\delta A^{\bullet M}\delta A^{\bullet J}}\right] \,.
    \label{eq:Eaction}
\end{multline}
Above, it is understood that the inverse propagator was factored out. 

One can see, from Fig.~\ref{fig:YM_oneloopEffaction}, that the effective action does consist of terms necessary to compute one-loop amplitudes. However, computing them using the effective action is not as straightforward as using the action itself. Therefore for the sake of completeness we recall the rules for computing amplitudes starting with the effective action in Appendix \ref{sec:AppA}. This will also be useful later in Section \ref{sec:Apps}.

\subsection{MHV action}
\label{subsec:MHV}

Before we derive the one-loop effective action for the MHV theory, let us briefly recall  it's structure at classical level. The MHV action is derived by applying the canonical field transformation to the Yang-Mills light cone action \eqref{eq:YM_LC_action} \cite{Mansfield2006}. In particular, the transformation is such that the self-dual part of the Yang-Mills action is mapped onto the free theory:
\begin{equation}
\mathrm{Tr}\,\hat{A}^{\bullet}\square\hat{A}^{\star}
+2ig\,\mathrm{Tr}\,\partial_{-}^{-1}\partial_{\bullet} \hat{A}^{\bullet}\left[\partial_{-}\hat{A}^{\star},\hat{A}^{\bullet}\right]
\,\, \longrightarrow \,\,
\mathrm{Tr}\,\hat{B}^{\bullet}\square\hat{B}^{\star}
\,,
\label{eq:MansfieldTransf1}
\end{equation}
where ${\hat B}^{\bullet}$ and ${\hat B}^{\star}$ are some new fields. The above requirement, together with the restriction that the transformation is canonical allows to obtain the expansions of the Yang-Mills fields in terms of the new fields. In momentum space, the solution for $\hat{A}^{\bullet}$ and $\hat{A}^{\star}$ read  \cite{Kotko2017}
\begin{equation}
    \widetilde{A}^{\bullet}_a(x^+;\mathbf{P}) = \sum_{n=1}^{\infty} 
    \int d^3\mathbf{p}_1\dots d^3\mathbf{p}_n \, \widetilde{\Psi}_n^{a\{b_1\dots b_n\}}(\mathbf{P};\{\mathbf{p}_1,\dots ,\mathbf{p}_n\}) \prod_{i=1}^n\widetilde{B}^{\bullet}_{b_i}(x^+;\mathbf{p}_i)\,,
    \label{eq:A_bull_exp}
\end{equation}
\begin{equation}
    \widetilde{A}^{\star}_a(x^+;\mathbf{P}) = \sum_{n=1}^{\infty} 
    \int d^3\mathbf{p}_1\dots d^3\mathbf{p}_n \, {\widetilde \Omega}_{n}^{a b_1 \left \{b_2 \cdots b_n \right \} }(\mathbf{P}; \mathbf{p_1} ,\left \{ \mathbf{p_2} , \dots ,\mathbf{p_n} \right \}) \widetilde{B}^{\star}_{b_1}(x^+;\mathbf{p}_1)\prod_{i=2}^n\widetilde{B}^{\bullet}_{b_i}(x^+;\mathbf{p}_i)\, ,
    \label{eq:A_star_exp}
\end{equation}
where ${\widetilde A}_a^{\bullet}$ and ${\widetilde A}_a^{\star}$ represent fields in momentum space at fixed light-cone time. The kernels read
\begin{equation}
    {\widetilde \Psi}_{n}^{a \left \{b_1 \cdots b_n \right \} }(\mathbf{P}; \left \{\mathbf{p}_{1},  \dots ,\mathbf{p}_{n} \right \}) =- (-g)^{n-1} \,\,  
    \frac{{\widetilde v}^{\star}_{(1 \cdots n)1}}{{\widetilde v}^{\star}_{1(1 \cdots n)}} \, 
    \frac{\delta^{3} (\mathbf{p}_{1} + \cdots +\mathbf{p}_{n} - \mathbf{P})\,\,  \mathrm{Tr} (t^{a} t^{b_{1}} \cdots t^{b_{n}})}{{\widetilde v}^{\star}_{21}{\widetilde v}^{\star}_{32} \cdots {\widetilde v}^{\star}_{n(n-1)}}  
      \, ,
    \label{eq:psi_kernel}
\end{equation}
\begin{equation}
    {\widetilde \Omega}_{n}^{a b_1 \left \{b_2 \cdots b_n \right \} }(\mathbf{P}; \mathbf{p}_{1} , \left \{ \mathbf{p}_{2} , \dots ,\mathbf{p}_{n} \right \} ) = n \left(\frac{p_1^+}{p_{1\cdots n}^+}\right)^2 {\widetilde \Psi}_{n}^{a b_1 \cdots b_n }(\mathbf{P};  \mathbf{p}_{1},  \dots , \mathbf{p}_{n}) \, .
    \label{eq:omega_kernel}
\end{equation}
Above, we use $p_1+\dots +p_n\equiv p_{1\dots n}$ and the curly braces represent the symmetrization with respect to the interchange of the color and momentum variables. Furthermore, the $\widetilde{v}_{ij}$, $\widetilde{v}^{\star}_{ij}$ are quantities similar to spinor products $\left<ij\right>$, $\left[ij\right]$, with the following explicit definitions:
\begin{equation}
    \widetilde{v}_{ij}=
    p_i^+\left(\frac{p_{j}^{\star}}{p_{j}^{+}}-\frac{p_{i}^{\star}}{p_{i}^{+}}\right), \qquad 
\widetilde{v}^*_{ij}=
    p_i^+\left(\frac{p_{j}^{\bullet}}{p_{j}^{+}}-\frac{p_{i}^{\bullet}}{p_{i}^{+}}\right)\, .
\label{eq:vtilde}
\end{equation}

\begin{figure}
    \centering
    \includegraphics[width=9cm]{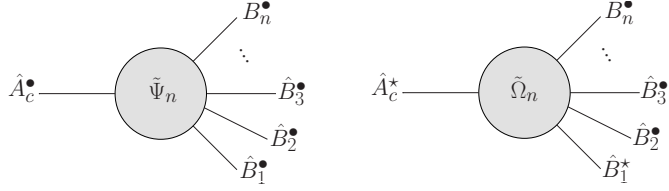}
    \caption{
    \small
    Graphical representation for the currents that give the Yang-Mills field expansion into the new fields $B$, see Eqs.~\eqref{eq:A_bull_exp}, \eqref{eq:A_star_exp}.
    }
    \label{fig:Psi_Omega}
\end{figure}

The kernels correspond to the Berends-Giele currents that encode a set of diagrams for on-shell external legs. It will be convenient to use diagrammatic representation for those currents, see Fig.~\ref{fig:Psi_Omega}.

Let us point out, that it is also possible to express the new fields in terms of the Yang-Mills fields. This is interesting, because it turns out that they are given by the straight infinite Wilson lines, integrated over all slopes \cite{Kotko2017,Kakkad2020} -- see Eq.~\eqref{eq:WilsonLineSol}.

Substituting the above solutions in Eq.~\eqref{eq:YM_LC_action} results in the MHV action 
\begin{equation}
S_{\mathrm{YM}}^{\left(\mathrm{LC}\right)}\left[{B}^{\bullet}, {B}^{\star}\right]=\int dx^{+}\left(
-\int d^{3}\mathbf{x}\,\mathrm{Tr}\,\hat{B}^{\bullet}\square\hat{B}^{\star} 
+\mathcal{L}_{--+}^{\left(\mathrm{LC}\right)}+\dots+\mathcal{L}_{--+\dots+}^{\left(\mathrm{LC}\right)}+\dots\right)\,,
\label{eq:MHV_action}
\end{equation}
where $\mathcal{L}_{--+\dots+}^{\left(\mathrm{LC}\right)}$ represents a generic $n$-point MHV vertex in the action, which in our conventions has the following form 
\begin{multline}
\mathcal{L}_{--+\dots+}^{\left(\mathrm{LC}\right)}=\int d^{3}\mathbf{p}_{1}\dots d^{3}\mathbf{p}_{n}\delta^{3}\left(\mathbf{p}_{1}+\dots+\mathbf{p}_{n}\right)\,
\widetilde{\mathcal{V}}_{--+\dots+}^{b_{1}\dots b_{n}}\left(\mathbf{p}_{1},\dots,\mathbf{p}_{n}\right)
\\ \widetilde{B}_{b_{1}}^{\star}\left(x^+;\mathbf{p}_{1}\right)\widetilde{B}_{b_{2}}^{\star}\left(x^+;\mathbf{p}_{2}\right)\widetilde{B}_{b_{3}}^{\bullet}\left(x^+;\mathbf{p}_{3}\right)\dots\widetilde{B}_{b_{n}}^{\bullet}\left(x^+;\mathbf{p}_{n}\right)
\,,
\label{eq:MHV_n_point}
\end{multline}
where 
\begin{equation}
\widetilde{\mathcal{V}}_{--+\dots+}^{b_{1}\dots b_{n}}\left(\mathbf{p}_{1},\dots,\mathbf{p}_{n}\right)= \!\!\sum_{\underset{\text{\scriptsize permutations}}{\text{noncyclic}}}
 \mathrm{Tr}\left(t^{b_1}\dots t^{b_n}\right)
 \mathcal{V}\left(1^-,2^-,3^+,\dots,n^+\right)
\,,
\label{eq:MHV_vertex}
\end{equation}
with the color ordered MHV vertex given by 
\begin{equation}
\mathcal{V}\left(1^-,2^-,3^+,\dots,n^+\right)= 
\frac{(-g)^{n-2}}{(n-2)!}  \left(\frac{p_{1}^{+}}{p_{2}^{+}}\right)^{2}
\frac{\widetilde{v}_{21}^{*4}}{\widetilde{v}_{1n}^{*}\widetilde{v}_{n\left(n-1\right)}^{*}\widetilde{v}_{\left(n-1\right)\left(n-2\right)}^{*}\dots\widetilde{v}_{21}^{*}}
\,.
\label{eq:MHV_vertex_colororder}
\end{equation}


\subsection{One-loop Effective MHV action}
\label{subsec:oneloopMHV}

\begin{figure}
    \centering
    \includegraphics[width=7cm]{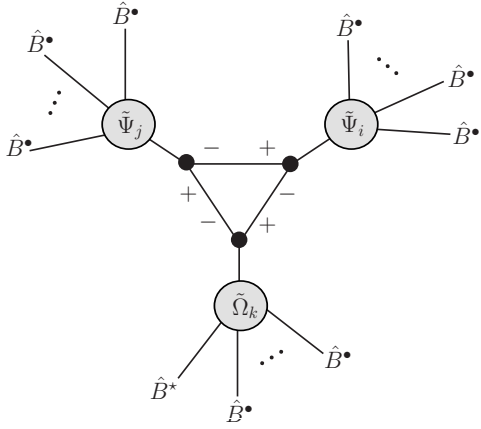}
    \caption{
    \small
    Expansion of the Yang-Mills fields in the $B$ fields in the loop terms of the effective action generates tree-level connections encoded in the field transformation kernels. }
    \label{fig:loop_3V}
\end{figure}

The idea is to start with the partition function for the light-cone Yang-Mills action Eq. \eqref{eq:Partition2}, which has all the necessary ingredients to obtain one-loop amplitudes, and apply the Mansfield's transformation to the classical fields:
\begin{equation}
\mathrm{Tr}\,\hat{A}^{\bullet}_c\square\hat{A}^{\star}_c
+2ig\,\mathrm{Tr}\,\partial_{-}^{-1}\partial_{\bullet} \hat{A}_c^{\bullet}\left[\partial_{-}\hat{A}_c^{\star},\hat{A}_c^{\bullet}\right]
\,\, \longrightarrow \,\,
\mathrm{Tr}\,\hat{B}_c^{\bullet}\square\hat{B}_c^{\star}
\,.\label{eq:MansfieldTransf2}
\end{equation}
The solutions to this transformation were given in the previous Section. Note, there is no Jacobian, even though it is trivial for the canonical transformation, as the field fluctuations around classical solutions have been already integrated out. Upon substitution, the classical light-cone Yang-Mills action becomes the MHV action, thus the partition function up to one loop becomes:
\begin{equation}
    Z_{\mathrm{MHV}}[J]\approx 
    \exp\left\{ iS_{\mathrm{MHV}}[B] 
    + i\int\!d^4x\, \Tr \hat{J}_i(x) \hat{A}_c^i[B](x) - \frac{1}{2} \Tr\ln \left( 
    \frac{\delta^2 S_{\mathrm{YM}}[A_c [B]]}
    {\delta \hat{A}^i(x)\delta \hat{A}^j(y)}
    \right)
    \right\}
    \label{eq:PartitionMHV}
    \,,
\end{equation}
Above, we explicitly denoted that the $\hat{A}_c$ fields in the two last terms are functionals of the ${\hat B}_c$ fields, $\hat{A}_c^i=\hat{A}_c^i[B]$. These are the inverse Wilson lines discussed earlier in Eqs.~\eqref{eq:A_bull_exp} and \eqref{eq:A_star_exp}. This has interesting consequences. Take for example the $(+ + -)$ triangle one-loop term generated by Eq.~\eqref{eq:Eaction} (there are two contributing diagrams with different helicity routing in the loop with two ${\hat A}^{\bullet}_c$ and one ${\hat A}^{\star}_c$ fields, see in Fig.~\ref{fig:YM_oneloopEffaction}). Consider the first term shown in the last line of Fig.~\ref{fig:YM_oneloopEffaction}. Each field ${\hat A}^i_c$ in this term will be replaced by inverse Wilson lines that can be expanded to any order after transformation.
This substitution takes care of all the tree-level connections that would develop in Yang-Mills if the $(+ + -)$ vertex in the action were combined with the loop terms in Eq.~\eqref{eq:Eaction} (use of the effective action to obtain scattering amplitudes is recalled in Appendix~\ref{sec:AppA}).  We show this pictorially in Fig.~\ref{fig:loop_3V}. Because the inverse Wilson lines already resum a subset of diagrams for $(+\dots +)$ and $(-+\dots +)$ helicity configurations, the number of contributions to one loop scattering amplitude in \eqref{eq:PartitionMHV} is reduced when compared with the number of terms obtained via the Yang-Mills partition function Eq.~\eqref{eq:Partition2}.

More crucially, when these solutions are substituted into the one-loop Yang-Mills effective action, in addition to the MHV vertices, we also obtain non-MHV vertices required at one loop.
Let us discuss this aspect in more details.

The terms originating from the Yang-Mills' kinetic portion $(+ -)$, upon substitution of the inverse Wilson lines Eq.~\eqref{eq:A_bull_exp}-\eqref{eq:A_star_exp}, are canceled out by the similar terms originating from the $(+ + -)$ vertex, leaving only the propagator (as shown in Eq.~\eqref{eq:MansfieldTransf2}). This cancellation, however, will not occur in the logarithm term in Eq.~\eqref{eq:PartitionMHV}, since 
it contains double differentiated action with respect to fields. Consequently, after the field substitution
\begin{equation}
 -\square_{IJ}-\left(V_{-++}\right)_{IJK}A_{c}^{\bullet K} \,\, \centernot\implies\,\,
-\square_{IJ} 
\,\,.
\label{eq:Mansfield2}
\end{equation}
The propagator term will not give any contributions that can cancel those created by the $(+ + -)$ vertex in the logarithm term. As a result, when the $\hat{A}_c^i[B(x)]$ is substituted, the vertex not only survives but also produces contributions. In other words,  after factoring out $\mathrm{det(\square)}$ as an infinite constant, the substitution of $\hat{A}_c^i=\hat{A}_c^i[B]$ will result in a non-zero loop contribution from the the self-dual part of the Yang-Mills action.
These contribution would be missing if one derived the one-loop effective action starting with the MHV action itself.

Let us illustrate the above considerations with an example. 
For simplicity, consider the one-loop two point $(+ -)$ amputated connected Green's function. We will not discard tadpoles in the argument below (we shall do so, however, in the practical computations later on). Expanding the logarithm term in Eq.~\eqref{eq:PartitionMHV} up to two point we get:
\begin{align}
   \frac{i}{2}\, \Bigg[2\, \square^{-1}_{IJ}\left(V_{-++}\right)_{JIK} B_c^{\bullet K} &+ 2\, \square^{-1}_{IJ}\left(V_{-++}\right)_{JIK}\Psi_2^{KK_1K_2} B_c^{\bullet K_1}B_c^{\bullet K_2}+2\, \square^{-1}_{IJ}\left(V_{-+-}\right)_{JIK} B_c^{\star K} \nonumber\\
   &+2\, \square^{-1}_{IJ}\left(V_{-+-}\right)_{JIK} \Omega_2^{KK_1K_2} B_c^{\star K_1}B_c^{\bullet K_2} +2\, \square^{-1}_{IJ}\left(V_{-++-}\right)_{JIKL} B_c^{\bullet K}B_c^{\star L} \nonumber\\
    &+\square^{-1}_{I_1J_1}\left(V_{-++}\right)_{J_1I_2K_1}
    \square^{-1}_{I_2J_2}\left(V_{-++}\right)_{J_2I_1K_2} 
    B_c^{\bullet K_1} B_c^{\bullet K_2}\nonumber\\
    &+\square^{-1}_{I_1J_1}\left(V_{-+-}\right)_{J_1I_2K_1}
    \square^{-1}_{I_2J_2}\left(V_{-+-}\right)_{J_2I_1K_2} 
    B_c^{\star K_1} B_c^{\star K_2}\nonumber\\
    &+\,\square^{-1}_{I_1J_1}\left(V_{-++}\right)_{J_1I_2K_1}
    \square^{-1}_{I_2J_2}\left(V_{-+-}\right)_{J_2I_1K_2} 
    B_c^{\bullet K_1} B_c^{\star K_2}\nonumber\\
    &+\,\square^{-1}_{I_1J_1}\left(V_{-+-}\right)_{J_1I_2K_1}
    \square^{-1}_{I_2J_2}\left(V_{-++}\right)_{J_2I_1K_2} 
    B_c^{\star K_1} B_c^{\bullet K_2}\nonumber\\
    &-\,\square^{-1}_{I_1J_1}\left(V_{++-}\right)_{J_1I_2K_1}
    \square^{-1}_{I_2J_2}\left(V_{--+}\right)_{J_2I_1K_2} 
    B_c^{\star K_1} B_c^{\bullet K_2}+ \dots \,\nonumber\\
    &-\,\frac{1}{2} \Big\{ 4\,\square^{-1}_{I_1J_1}\left(V_{-++}\right)_{J_1I_2K_1}
    \square^{-1}_{I_2J_2}\left(V_{-++}\right)_{J_2I_1K_2} 
    B_c^{\bullet K_1} B_c^{\bullet K_2} \nonumber\\
    &+4\,\square^{-1}_{I_1J_1}\left(V_{-+-}\right)_{J_1I_2K_1}
    \square^{-1}_{I_2J_2}\left(V_{-+-}\right)_{J_2I_1K_2} 
    B_c^{\star K_1} B_c^{\star K_2}\nonumber\\
    &+4\,\,\square^{-1}_{I_1J_1}\left(V_{-++}\right)_{J_1I_2K_1}
    \square^{-1}_{I_2J_2}\left(V_{-+-}\right)_{J_2I_1K_2} 
    B_c^{\bullet K_1} B_c^{\star K_2}\nonumber\\
    &+4\,\,\square^{-1}_{I_1J_1}\left(V_{-+-}\right)_{J_1I_2K_1}
    \square^{-1}_{I_2J_2}\left(V_{-++}\right)_{J_2I_1K_2} 
    B_c^{\star K_1} B_c^{\bullet K_2} + \dots \, \Big\}
    \Bigg]\,,
    \label{eq:2p_1loopexpMHV}
\end{align}
where $\Psi_2^{KK_1\dots K_n}$ and $\Omega_2^{KK_1\dots K_n}$ represent position space version of the kernels Eq.~\eqref{eq:psi_kernel}, \eqref{eq:omega_kernel} respectively, written in the collective index notation.  The terms in the expression Eq.~\eqref{eq:2p_1loopexpMHV} above are essentially those in Eq.~\eqref{eq:2p_1loopexp} with the $\hat{A}_c^i[B]$ fields expanded up to second order.
The rules of how to use the effective action to calculate amplitude are reminded in Appendix~\ref{sec:AppA}.
For the two-point case, however, it is simpler to expand back the classical fields $B_c$ into external sources $J$.  
There are two kinds of contributions:
\begin{enumerate}[label={\it\roman*}$\,$)]
   \item  The first set of contributions comes from the first order expansion of $B$ fields into sources, which is equivalent to 
   \begin{equation}
     \frac{i}{2}\Bigg[\frac{\delta^2}
    {\delta B_c^{\bullet K_1} \delta B_c^{\star K_2}} \left\{ \Tr\ln \left( 
    \frac{\delta^2 S_{\mathrm{YM}}[A_c [B]]}
    {\delta \hat{A}^i(x)\delta \hat{A}^j(y)}
    \right)
    \right\}\Bigg]_{B_c= 0}\,.
   \end{equation}
   Substituting Eq.~\eqref{eq:2p_1loopexpMHV} we get:
   \begin{multline}
   \frac{i}{2}\, \Bigg[  2\, \square^{-1}_{IJ}\left(V_{-+-}\right)_{JIK} \Omega_2^{KK_2K_1}  +2\, \square^{-1}_{IJ}\left(V_{-++-}\right)_{JIK_1K_2} \\
   +\,\square^{-1}_{I_1J_1}\left(V_{-++}\right)_{J_1I_2K_1}
    \square^{-1}_{I_2J_2}\left(V_{-+-}\right)_{J_2I_1K_2} \\
    +\,\square^{-1}_{I_1J_1}\left(V_{-+-}\right)_{J_1I_2K_2}
    \square^{-1}_{I_2J_2}\left(V_{-++}\right)_{J_2I_1K_1} \\
    -\,\square^{-1}_{I_1J_1}\left(V_{++-}\right)_{J_1I_2K_2}
    \square^{-1}_{I_2J_2}\left(V_{--+}\right)_{J_2I_1K_1} \\
    -\,\frac{1}{2} \Big\{ 4\,\,\square^{-1}_{I_1J_1}\left(V_{-++}\right)_{J_1I_2K_1}
    \square^{-1}_{I_2J_2}\left(V_{-+-}\right)_{J_2I_1K_2} \\
    +4\,\,\square^{-1}_{I_1J_1}\left(V_{-+-}\right)_{J_1I_2K_2}
    \square^{-1}_{I_2J_2}\left(V_{-++}\right)_{J_2I_1K_1}  \, \Big\}
    \Bigg]\,.
    \label{eq:2p_loopmhv}
\end{multline}
In terms of diagrams the contributing terms are: 
\begin{center}
\includegraphics[width=11cm]{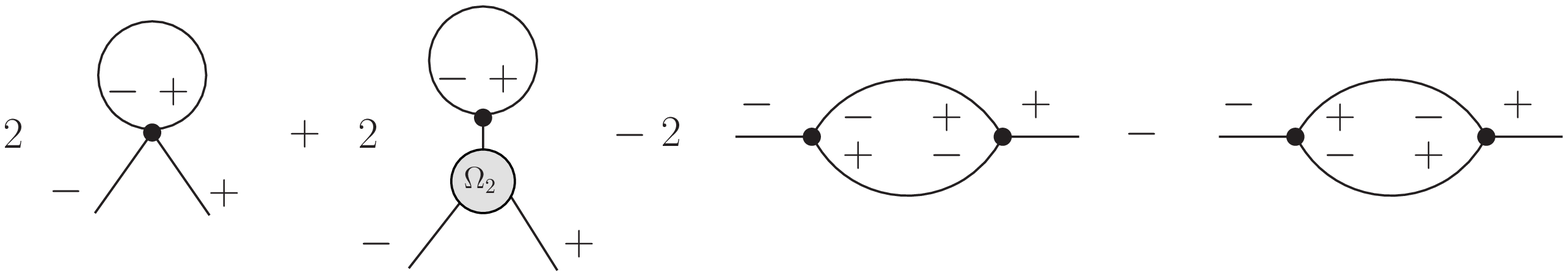}
\label{fig:X}
\end{center}
   \item 
   The tree-level connection between the classical $S_{\mathrm{MHV}}[B]$ and the logarithm term. This contribution appears due to expansion of the $B_c$ fields in the external sources to second order. There is only one contribution of this type to the one-loop $(+ -)$ case
   \begin{equation}
        -\frac{i}{2}\, \Bigg[2\, \square^{-1}_{IJ}\left(V_{-++}\right)_{JIL} \,\square^{-1}_{LK}\left(V^{\mathrm{MHV}}_{--+}\right)_{KK_1K_2}\,\Bigg]\,.
   \end{equation}
   This corresponds to the following diagram:
   \begin{center}
\includegraphics[width=2.5cm]{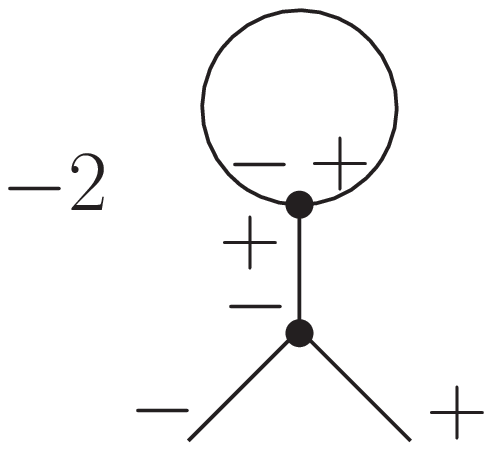}
\label{fig:Y}
\end{center}
\end{enumerate}
\begin{figure}
    \centering
    \includegraphics[width=13cm]{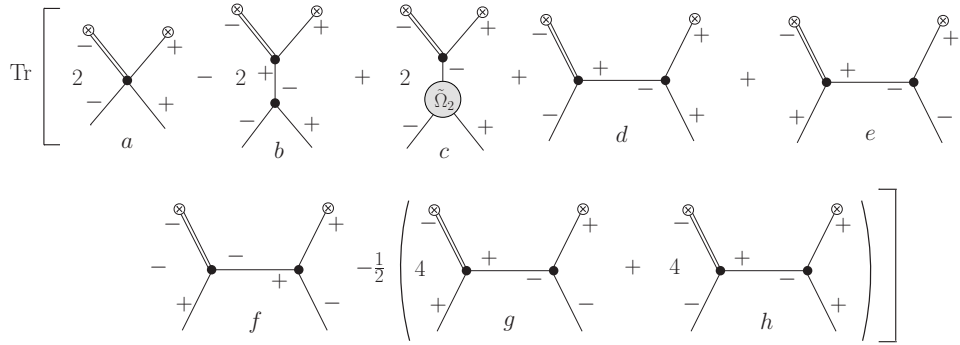}
    \caption{
    \small 
    Diagrams contributing to one-loop $(+-)$ self energy diagram. For the discussion purpose, we show the diagrams as un-traced, i.e. the loop diagrams result in connecting the x-circles. The double line indicates a propagator.
    }
    \label{fig:2point}
\end{figure}
Now, in order to show that the MHV vertices appear inside the loops, let us un-trace the above contributions, i.e. cut open the loops -- see Fig.~\ref{fig:2point}. These un-traced diagrams become the normal loop diagrams, shown above, after connecting the crossed circles, which represent the differentiated leg of the action, cf. Eq.~\eqref{eq:PartitionMHV}. Note, that one of the differentiated legs contains the propagator (denoted as double line).
Contributions b) and c) arise from the $(+ + -)$ and $(+ - -)$ un-traced tadpole terms. The former involves tree level connection with the 3 point MHV vertex in $S_{\mathrm{MHV}}[B]$ and the latter involves expansion of $\hat{A}_c^{\star}[B]$ to second order. It turns out, that the sum of terms a), b), c) and h) is equal to 4-point un-traced MHV, see Fig.~\ref{fig:4pointMHV}. The details of this identity are discussed in Appendix~\ref{sec:AppB}. 
\begin{figure}
    \centering
    \includegraphics[width=13cm]{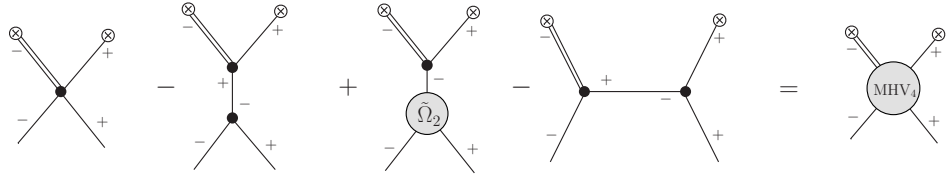}
    \caption{
    \small
    Combination of un-traced diagrams, that leads to the MHV vertex under the logarithm term in the one-loop effective action.
    }
    \label{fig:4pointMHV}
\end{figure}
We have thus generated a loop diagram with the MHV vertex inside a loop. 
This is precisely the only diagram one would obtain from the classical MHV action, applying the standard Feynman rules.
However, 
we see, that there are diagrams left, namely contributions d), e), f) and g). 
These are precisely the terms missing in the MHV action. Notice, that these contribution arise from the inter mixing of the self-dual part $(+ + -)$ with the 3 point MHV vertex $(+ - -)$. 

\begin{figure}
    \centering
    \includegraphics[width=10cm]{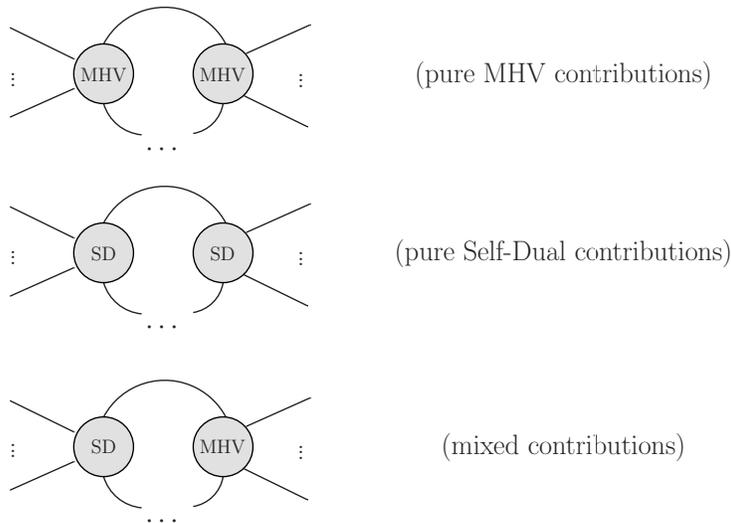}
    \caption{
    \small
    Types of one-loop diagrams originating in the effective action for the MHV theory. The first class (top) are diagrams that involve only MHV vertices in the loop. These are the diagrams that would follow from classical MHV action. In addition to that, the one-loop effective action generates diagrams which involve pure Self-Dual interactions in the loop (middle), as well as the mixed contributions (bottom).
    }
    \label{fig:MHVloop_generic}
\end{figure}

The result can be further generalized. MHV action misses one-loop contributions that originate either from the self-dual part of the Yang-Mills action alone or from the inter-mixing of the self-dual part $(+ + -)$ with the MHV vertices. 
This is summarized in Fig.~\ref{fig:MHVloop_generic}.

In the following section we use Eq.~\eqref{eq:PartitionMHV} to explicitly compute some pure gluonic one-loop amplitudes.
\section{Applications}
\label{sec:Apps}

An example of an amplitude that cannot be computed with just the MHV vertices (and thus the classical MHV action) are one-loop all-plus helicity amplitudes $(+\dots +)$ \cite{Brandhuber2007a,Fu_2009, Boels_2008, Ettle2007,Brandhuber2007,Elvang_2012}. This amplitude is zero at the tree level but non-zero in general \cite{Bern_1994}. At one loop it
is given by a rational function of the spinor products, which, in our conventions  reads (modulo normalization):
   \begin{equation}
       \mathcal{A}^{\mathrm{one-loop}}_n(+ + \dots +) = g^n\sum_{1 \leq i < j < k < l \leq n} \frac{\widetilde{v}_{ij}^{*}\widetilde{v}_{jk}\widetilde{v}_{kl}^{*}\widetilde{v}_{li}}{\widetilde{v}_{1n}^{*}\widetilde{v}_{n\left(n-1\right)}^{*}\widetilde{v}_{\left(n-1\right)\left(n-2\right)}^{*}\dots\widetilde{v}_{21}^{*}} \, .
       \label{eq:all_plus_one_loop}
   \end{equation}
Similarly, $(+ + \dots + -)$ one-loop helicity configuration cannot be calculated from the classical MHV action and similarly is a rational function of spinor products.

In order to validate our quantum MHV action and demonstrate how it is applied, we compute these two amplitudes for the four point case. 
Our strategy is the following:
\begin{enumerate}[label={\it\roman*}$\,$)]
\item determine the complete one-loop contributions using Eq.~\eqref{eq:PartitionMHV} and rules from Appendix~\ref{sec:AppA}; these contributions will contain the inverse Wilson line kernels that include tree-level self-dual interactions,
\item determine the one-loop bubble, triangle and box off-shell sub-diagrams in the 4D world-sheet regularization scheme using calculations of \cite{CQT1,CQT2}; we recalculate some of the sub-diagrams in Appendix~\ref{sec:App_CQT_calculations},
\item convolute the sub-diagrams with the inverse Wilson line kernels to obtain the full amplitude.
\end{enumerate}

Before we discuss the amplitudes, in the next Section we motivate and briefly review the world-sheet regularization scheme.

\subsection{The CQT world-sheet regularization}
\label{sub:CQT}

In order to preserve the simple form of the MHV vertices it is crucial to work in four space-time dimensions. 
As discussed in \cite{Ettle2007}, the dimensionally regulated MHV theory does not preserve the simple structure of the MHV vertices Eq.~\eqref{eq:MHV_vertex_colororder}; indeed, the single-term holomorphic MHV vertices are inherently a feature of the four dimensional space.  In addition, the compact holomorphic form for the kernels Eqs.~\eqref{eq:A_bull_exp} and \eqref{eq:A_star_exp} is no longer preserved. Therefore to regularize the divergent integrals we use the world-sheet regularization of Chakrabarti, Qiu, and Thorn (CQT)  \cite{CQT1,CQT2} which is entirely four dimensional scheme. 

A critical element of this technique is that 
the loop integrals are expressed in terms of the so-called "region momenta", instead of the line momenta, which are dual in the sense that they provide the world-sheet description of the planar scattering process \cite{HOOFT1974461,Bardakci_2002}. Although, originally the motivation of this technique was to study the connection between string theories and quantum field theories, it can be also used as a computational scheme in light cone variables in four dimensions. 

\begin{figure}
    \centering
    \includegraphics[width=6cm]{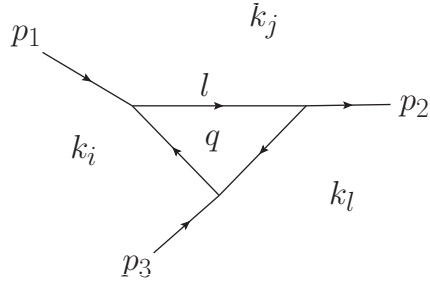}
    \caption{An example diagram illustrating the "region momenta" assignment. The diagram divides the plane into regions; the momentum associated with the confined region enclosed by the loop is $q$, whereas for the regions extending to infinity it is $k_i$. The line momenta are $p_i$ for the external lines and $l$ for the loop momentum. The arrows denote the orientation of the line momenta.
    }
    \label{fig:reg_mom}
\end{figure}

Consider for example a one loop planar diagram shown in Fig. \ref{fig:reg_mom}. The plane gets divided into regions by the internal and external lines. External lines bound regions stretching to infinity, while loops bound confined regions in the plane. In the world-sheet representation of planar diagrams one assigns the region (or dual) momentum variables $q_i$ for the finite regions and $k_j$ for the exterior regions. The difference of the dual momentum variables of the two regions represents the actual momentum carried by the line separating them. We  adopt the following convention:
\begin{equation}
    \mathrm{Line\,\, Momentum} =  \mathrm{Region\,\, Momentum\,\,on\,\, right} -  \mathrm{Region\,\, Momentum\,\,on\,\, left}\,,
    \label{eq:LM_con}
\end{equation}
where, "right" and "left" are defined with respect to the direction of the line momentum.
Thus for the line momenta shown in Fig.~\ref{fig:reg_mom}, we get:
\begin{equation}
    p_1= k_i-k_j \,,\quad p_2= k_l-k_j\,, \quad p_3= k_l-k_i\,, \quad l=q-k_j \,.
    \label{eq:line_mom}
\end{equation}
The above mapping converts the loop integration to integration over the dual momentum $q$. 

The region momenta prescription is, undeniably, applicable solely to planar diagrams. This, however, is adequate because in the present work only the leading single-trace part of the one-loop amplitude is calculated.

The CQT approach uses a simple exponential factor to cutoff integrals in the transverse directions and a discretization of the "plus" light cone component of the dual momentum. 
Using the notation considered above, for a generic
$n$-loop diagram, one inserts:
\begin{equation}
    \exp \Big( -\delta \sum_{i=1}^n \mathbf{q}_i^2 \Big) \,
    \label{eq:cqt_exp}
\end{equation}
in the loop integrand. The parameter
$\delta$ is positive, and in the end it is put to zero. Since,
$\mathbf{q}_i^2 = 2q^{\bullet}q^{\star}$ involves only the transverse components of the loop dual-momenta, it explicitly violates Lorentz invariance. This results in non-zero values for certain 
processes violating Lorentz invariance, which in turn must be canceled by introducing appropriate counterterms. These counterterms, and also the ones used
to cancel  divergences can be systematically introduced in the  effective action. 

Consider for instance a generic one-loop counterterm in the CQT scheme. The exact one-loop counterterms in this scheme are discussed in the following sub-sections where we consider the one-loop amplitudes. These counterterms can be introduces in the Yang-Mills partition function \eqref{eq:Partition2} as follows: 
\begin{equation}
    Z_{\mathrm{YM}}[J]\approx 
    \exp\left\{ iS_{\mathrm{YM}}[A_c] 
    + i\int\!d^4x\, \Tr \hat{J}_i(x) \hat{A}_c^i(x) - \frac{1}{2} \Tr\ln \left( 
    \frac{\delta^2 S_{\mathrm{YM}}[A_c]}
    {\delta \hat{A}^i(x)\delta \hat{A}^j(y)}
    \right) + i\Delta S_\mathrm{YM}^\mathrm{CQT}[A_c]
    \right\}
    \label{eq:PartitionYM_CT}
    \,,
\end{equation}
where $\Delta S_\mathrm{YM}^\mathrm{CQT}$ are counterterms.
Applying the Mansfield's transformation that maps the Yang-Mills theory to the MHV theory we get:
\begin{multline}
    Z_{\mathrm{MHV}}[J]\approx 
    \exp\Bigg\{ iS_{\mathrm{MHV}}[B] 
    + i\int\!d^4x\, \Tr \hat{J}_i(x) \hat{A}_c^i[B](x) - \frac{1}{2} \Tr\ln \left( 
    \frac{\delta^2 S_{\mathrm{YM}}[A_c [B]]}
    {\delta \hat{A}^i(x)\delta \hat{A}^j(y)}
    \right)\\
    + i\Delta S_\mathrm{YM}^\mathrm{CQT}[A_c[B]]
    \Bigg\}
    \label{eq:PartitionMHV_CT}
    \,,
\end{multline}
where $\Delta S_\mathrm{YM}^\mathrm{CQT}[A_c[B]]$ represents a series of counterterms originating via the substitution of the inverse Wilson lines Eq.~\eqref{eq:A_bull_exp}-\eqref{eq:A_star_exp}. These terms explicitly cancel similar contributions coming from the logarithm term.

In the following sub-sections we consider $(++++)$ and $(+++-)$ one loop amplitudes using  Eq.~\eqref{eq:PartitionMHV}.

\subsection{$(+ + + +)$ one-loop amplitude.}
\label{sub:4plus}

\begin{figure}
    \centering
    \includegraphics[width=15.6cm]{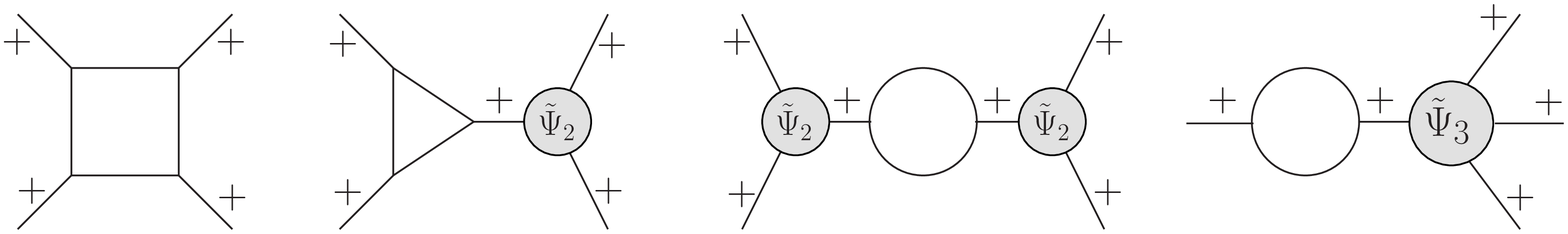}
    \caption{
    The first three diagrams from the left represent types of contributions to the $(+ + + +)$ one-loop amplitude obtained from Eq.~\eqref{eq:PartitionMHV}. Inclusion of the last diagram (with external bubble) makes the sum of all diagrams zero, what is useful to evaluate amplitude -- see the main text.
    Notice, that there is no propagator between the $\Psi_n$ kernels and the rest of the diagram -- it is included inside the kernel, which resums the tree-level connections.
}
    \label{fig:4plus_EA}
\end{figure}

As discussed, the all-plus helicity one-loop amplitude is one of the amplitudes which cannot be computed from the classical MHV action given by Eq.~\eqref{eq:MHV_action}. This is because it
gets contribution solely from the self-dual part of the Yang-Mills action Eq.~\eqref{eq:YM_LC_action}, which via the Mansfield's transformation has been mapped to a free theory.
The contributions to the all-plus amplitude arise only from the logarithm term in Eq.~\eqref{eq:PartitionMHV}. Using the rules reviewed in Appendix~\ref{sec:AppA}, we obtain the three first terms depicted in Fig.~\ref{fig:4plus_EA} (we drop the symmetry factors here for brevity). Interestingly, when diagrams with bubble on external leg (the last diagram) are included with appropriate symmetry factor the sum turns out to be zero. Thus, one can use the bubble diagrams, which in CQT scheme have very simple form, to obtain the triangle and box diagrams. This identity has been demonstrated in \cite{CQT1} for ordinary diagrams; here we repeated this for  the diagrams from Fig.~\ref{fig:4plus_EA} that contain the inverse Wilson line kernels $\Psi_2$, $\Psi_3$ and got the same result. Indeed, those kernels just resum the tree-level connections and include an appropriate propagator.

The helicity non-conserving gluon self energy  $(++)$ does not vanish in CQT scheme even on-shell. It is a consequence of worldsheet-friendly but Lorentz-violating regularization scheme.
It turns out to be a quadratic polynomial in the dual momenta \cite{CQT1} (see also Appendix~\ref{sec:App_CQT_calculations}):
\begin{equation}
    \Pi^{+ +}= \frac{g^{2}}{12 \pi^{2}} \left[k_{0}^{\star 2}+k_{1}^{\star 2}+k_{0}^{\star} k_{1}^{\star}\right]\,.
    \label{eq:++GSE}
\end{equation}
According to the CQT procedure, such a term has to be removed from the action by an explicit counterterm: 
\begin{center}
 \includegraphics[width=11cm]{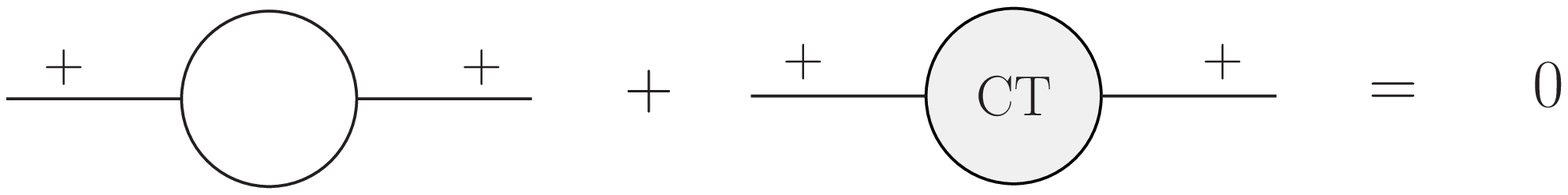}
\end{center}
generated by the $\Delta S_\mathrm{YM}^\mathrm{CQT}[A_c[B]]$ terms in \eqref{eq:PartitionMHV_CT}.

Once the counterterms have been introduced, we have therefore only the initial two contribution illustrated in Fig.~\ref{fig:4plus_EA}, which are equal to the (negative) of the sum of last two diagrams. 
Both ways can be used to obtain the amplitude. Let us focus on evaluating the first two diagrams in the CQT scheme.
The box diagram and the triangle sub-diagram have been calculated in \cite{CQT1}. In Appendix~\ref{sec:App_CQT_calculations} we review some of the calculations and cite necessary formulae. Here we discuss the assembly of the final result.

\begin{figure}
    \centering
    \includegraphics[width=15.6cm]{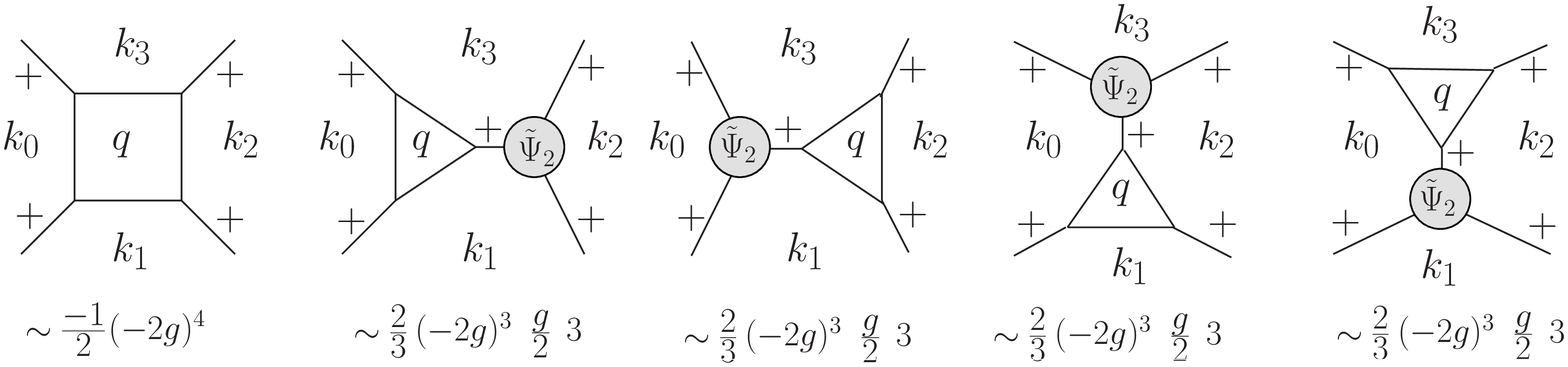}
    \caption{The contributing terms to the all-plus $(+ + + +)$ helicity one-loop amplitude with explicit region momenta assignments and symmetry factors. $q$ is the region momentum associated with the loop and $k_i$ with the external region. ${\tilde \Psi}_2$ represents the inverse Wilson line kernel Eq.~\eqref{eq:psi_kernel} expanded to second order.}
    \label{fig:all_plus1}
\end{figure}

The exact diagrams and the associated symmetry factors are shown in Fig.~\ref{fig:all_plus1}. Each 3-point vertex in momentum space has an associated factor of $(-2g)$ (see Eq.~\eqref{eq:v3_position}). The additional $-1/2$ for the box contribution $\square^{+ + + +}$ comes from the sum of all such contributions acquired by expanding the logarithm term  to four points. The symmetry factor for the triangular graph via the logarithmic expansion  is $2/3$. The $g/2$ is from the second-order expansion of the inverse Wilson line Eq.~\eqref{eq:A_bull_exp} substituted to a given leg. The other two legs contribute similarly, resulting in the overall factor of $3$. 

The triangular contributions in Fig.~\ref{fig:all_plus1}, which we denote as $\Delta_{ij}^{+ + + +}$ with $i$,$j$ external on-shell legs in a triangle, have been re-derived in Appendix~\ref{sec:App_CQT_calculations}, following \cite{CQT1}. They read:
\begin{equation}
   \Delta_{12}^{+ + + +}= \frac{-g^{4}}{12 \pi^{2}} \frac{\left({\tilde v}_{12}p_2^+\right)^{3}p_{3}^{+}}{p_{1}^{+} p_{2}^{+} p_{3}^{+}p_{4}^{+} p_{34}^{2}{\tilde v}^*_{34}}\,,
\end{equation}
\begin{equation}
    \Delta_{41}^{+ + + +}= \frac{-g^{4}}{12 \pi^{2}} \frac{\left({\tilde v}_{41}p_1^+\right)^{3}p_{2}^{+}}{p_{1}^{+} p_{2}^{+} p_{3}^{+} p_{4}^{+} p_{14}^{2}{\tilde v}^*_{23}}\,,
\end{equation}
\begin{equation}
   \Delta_{23}^{+ + + +}= \frac{-g^{4}}{12 \pi^{2}} \frac{ \left({\tilde v}_{23}p_3^+\right)^{3}p_{4}^{+}}{p_{1}^{+} p_{2}^{+} p_{3}^{+} p_{4}^{+} p_{14}^{2}{\tilde v}^*_{41}}\,.
\end{equation} 
It turns out, that the
 box diagram $\square^{+ + + +}$ can be reduced to the triangular contributions  \cite{CQT1}. 
For the box term in Fig.~\ref{fig:all_plus1} this reduction gives $-\Delta_{34}^{+ + + +}$, which essentially is the negative of the last term
in Fig.~\ref{fig:all_plus1}, plus additional contributions. The sum of both diagrams read
\begin{multline}
  \square^{+ + + +} \,+\, \Delta_{34}^{+ + + +} = \frac{-g^{4}}{12 \pi^{2}} \frac{p_{1}^{+}}{p_{1}^{+} p_{2}^{+} p_{3}^{+} p_{4}^{+} {\tilde v}^*_{12} p_{14}^{2}}\\
  \left[{\tilde v}_{41}p_{1}^{+} {\tilde v}_{23}p_{3}^{+}\left({\tilde v}_{41}p_{1}^{+}+{\tilde v}_{23}p_{3}^{+}\right)+{\tilde v}_{34}p_{4}^{+}\left({\tilde v}_{41}^{2}p_{1}^{+ 2}+{\tilde v}_{23}^{2}p_{3}^{+ 2}\right)\right]\,.
    \label{eq:box+tri}
\end{multline}

With a bit of algebra, one can show that the sum of all contributions in the on-shell limit is
\begin{equation}
      \mathcal{A}_{\mathrm{one-loop}}^{+ + + +} = \frac{g^{4}}{24 \pi^{2}} \frac{{\tilde v}_{21} {\tilde v}_{43}}{{\tilde v}^{\star}_{21} {\tilde v}^{\star}_{43}}     \, .
       \label{eq:4_plus_one_loop}
\end{equation}
The above result is consistent with the known result \cite{Bern:1991aq,Kunszt_1994}, Eq.~\eqref{eq:all_plus_one_loop} (modulo a normalization).

\subsection{$(+ + + -)$ one-loop amplitude.}
\label{sub:3plus_minus}

\begin{figure}
    \centering
 \includegraphics[width=13cm]{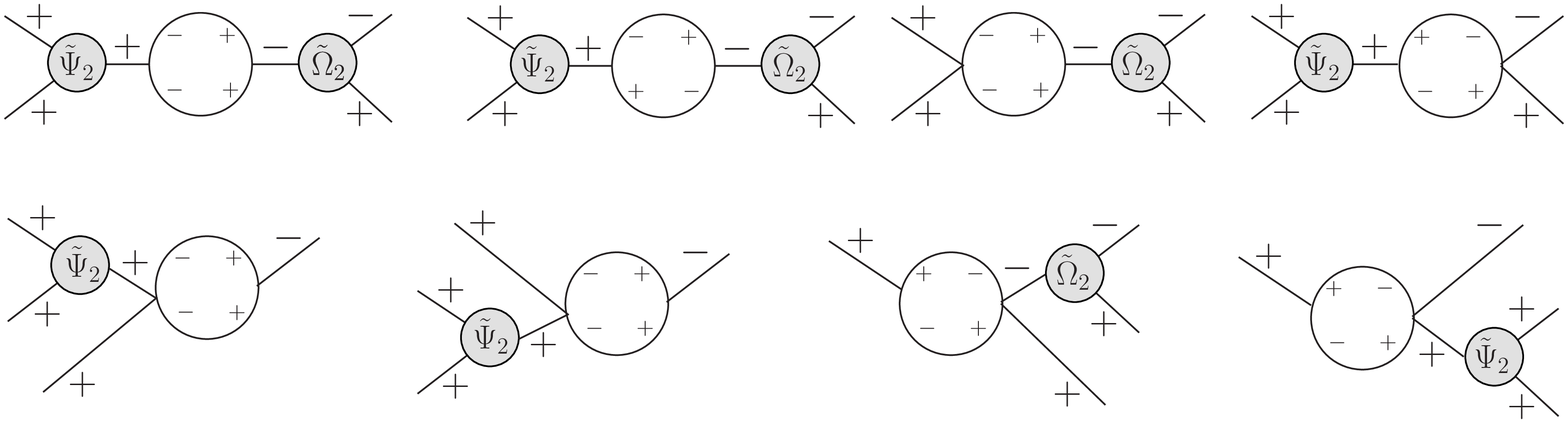}\\
 \includegraphics[width=13cm]{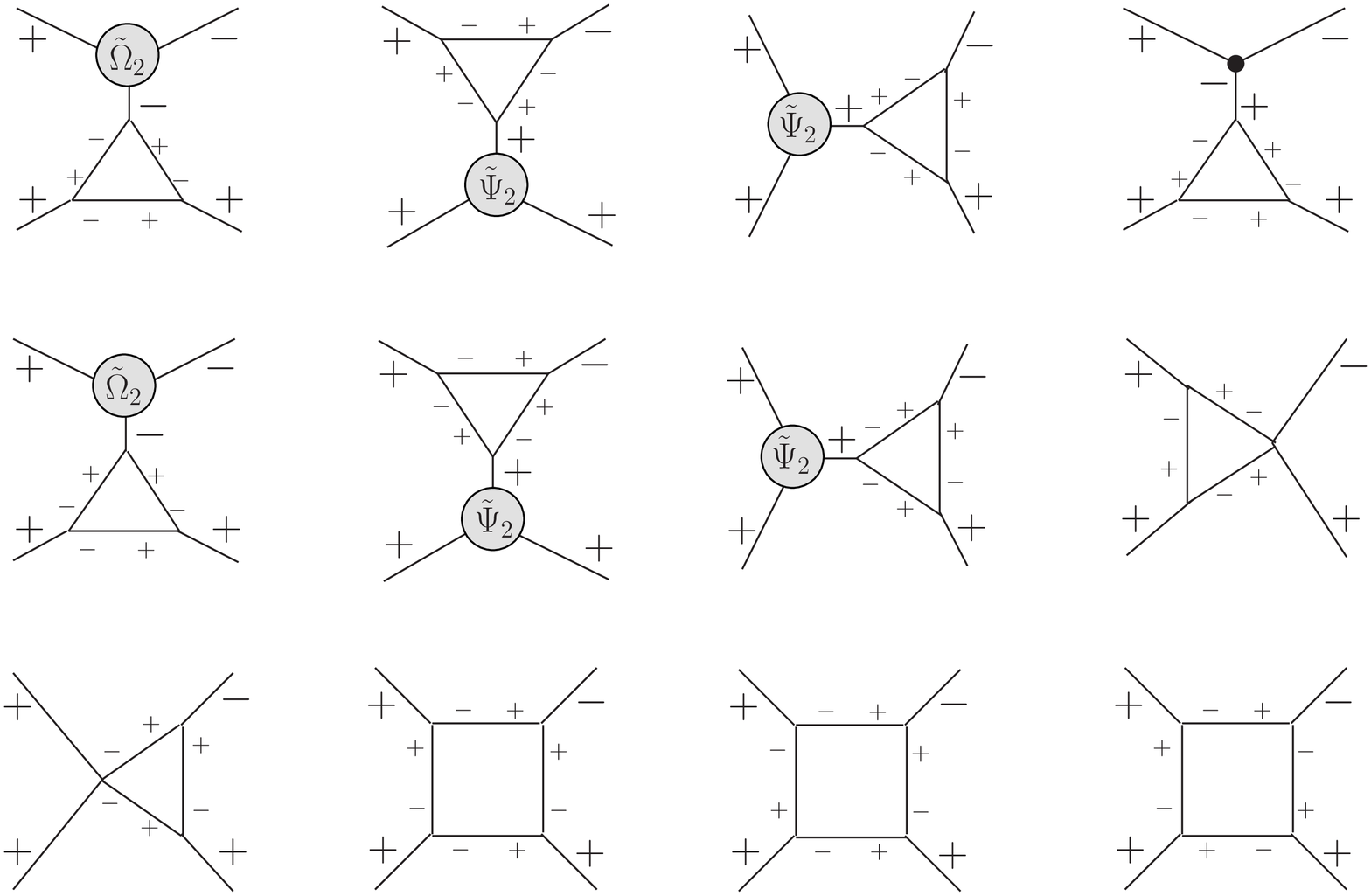} 
    \caption{\small The contributions to $(+ + + -)$ one-loop amplitude originating from \eqref{eq:PartitionMHV} via the rules discussed in Appendix \ref{sec:AppA}. For simplicity we suppressed the symmetry factors and the line momenta associated with each term. The kernels ${\tilde \Psi}_2$ and  ${\tilde \Omega}_2$ result from the substitution of inverse Wilson line Eq.~\eqref{eq:A_bull_exp}-\eqref{eq:A_star_exp}. The $\bullet$ in the last term of line 3 represents the $(+ - -)$ three point MHV vertex connected with the $\Delta^{+ + +}$ via a propagator.}
    \label{fig:3plusminus}
\end{figure}

The contributions generated from the effective action Eq.~\eqref{eq:PartitionMHV} 
are shown in Fig.~\ref{fig:3plusminus}. 
We name the legs $+ + + -$, respectively, as 1, 2 , 3  and 4 with corresponding line momenta ${p}_{i}$ anti-clockwise.
Furthermore, we consider the color ordered case and suppress the color factors for simplicity. These diagrams, unlike in the previous case,  are  both ultraviolet and infrared divergent. The strategy we use to compute the amplitude is the following. 
The one-loop 1PI contributions coming from the log term in Eq.~\eqref{eq:PartitionMHV}, prior to the inverse Wilson line substitution, are determined using the explicit calculation done in \cite{CQTnotes,CQT1}, with appropriate symmetry factors.
These are then convoluted with the inverse Wilson line kernels Eq.~\eqref{eq:A_bull_exp},\eqref{eq:A_star_exp}. Below, we describe how the final result is assembled, quoting necessary results from \cite{CQT1}.

The $(+ +)$ gluon self-energy, as stated previously, is non-zero and will be explicitly canceled by an appropriated counter term.
The $(+ -)$ gluon self-energy is calculated in the same way as the $(+ +)$ gluon self-energy  (see Appendix~\ref{sec:App_CQT_calculations} for a review of CQT scheme calculations).
In this case we get terms with two types of divergences.
A quadratic divergence, manifesting itself as a pole  $1/\delta$, and a logarithmic divergence. The procedure of dealing with these divergence was described in detail in \cite{CQT1}. 
The result reads \cite{CQT1}:
\begin{equation}
\Pi^{+ -} =\frac{g^{2}}{4 \pi^{2}} p^{2}\left(\sum_{q^{+}}\left[\frac{1}{q^{+}}+\frac{1}{p^{+}-q^{+}}\right] \ln \left\{\frac{q^{+}\left(p^{+}-q^{+}\right)}{p^{+2}} p^{2} \delta e^{\gamma}\right\}-\frac{11}{6} \ln \left(p^{2} \delta e^{\gamma}\right)+\frac{67}{18}\right)\,,
\label{eg:pi+-}
\end{equation}
where $\gamma$ is the Euler's constant. Above, $q$ is the region momenta associated with the loop and $p$ is the external line momentum. 
In CQT scheme, 
the $q^+=0$ singularity is regulated by the discretization procedure, as each world-sheet is parametrized by the momentum plus component. In discretized space $q^+ = l P^+$, where $l=1,2,\dots,N$, with  $N$ being the discretized total plus momentum entering the self-energy diagram $p^+ = N P^+$ (here $P^+$ sets the mass scale). Thus, the summation over the discrete momentum is defined as
$\sum_{q^{+}}= P^+\sum_{l=1}^{N-1}$. The $q^+=0$  divergence is related to the light-cone gauge choice and will be explicitly canceled by similar contributions arising from other diagrams.
Substituting the inverse Wilson lines Eq.~\eqref{eq:A_bull_exp},\eqref{eq:A_star_exp} to the $(+-)$ gluon self-energy gives rise to the first two terms in Fig.~\ref{fig:3plusminus}. These contributions, with appropriate symmetry factors, arising via the rules discussed in Appendix \ref{sec:AppA}, read: 
\begin{multline}
\mathcal{A}_{\mathrm{SE}}^{+ + + -} =\frac{ -g^{4}p_{4}^{+}}{2 \pi^{2} p_{1}^{+} p_{2}^{+} p_{3}^{+}}\Bigg[\frac{{\tilde v}_{34}p_{4}^{+}p_{2}^{+}}{{\tilde v}_{21}^*}\\
\Bigg\{\sum_{q^{+}}\left[\frac{1}{q^{+}}+\frac{1}{p_{12}^{+}-q^{+}}\right] \ln \Bigg(\frac{q^{+}\left(p_{12}^{+}-q^{+}\right)}{p_{12}^{+ 2}} p_{12}^{2} e^{\gamma} \delta\Bigg)
-\frac{11}{6} \ln \left(p_{12}^{2} e^{\gamma} \delta\right)+\frac{67}{18}\Bigg\}\\
+\frac{{\tilde v}_{41}p_{3}^{+}p_{1}^{+}}{{\tilde v}_{32}^*}\Bigg\{\sum_{q^{+}+p_4^+}\left[\frac{1}{q^{+}+p_4^+}+\frac{1}{p_1^{+}-q^{+}}\right] \ln \Bigg(\frac{(q^{+}+p_4^+)\left(p_1^{+}-q^{+}\right)}{p_{14}^{+2}} p_{14}^{2} e^{\gamma} \delta\Bigg)\\
-\frac{11}{6} \ln \left(p_{14}^{2} e^{\gamma} \delta\right)+\frac{67}{18}\Bigg\}\Bigg] \,.
\label{eq:self_contri_OL}
\end{multline}
Above, the subscript $\mathrm{SE}$ stands for "Self-Energy". 

Next, we consider the diagrams involving the triangle and sword fish diagrams. In Fig.~\ref{fig:3plusminus}, the last two diagrams in line 1 and the whole  line 2 represents the sword fish terms, whereas the line 3 and first 3 diagrams in line 4  represents the triangular contributions.
For triangle diagrams, the contributing diagrams consist in the one-loop triangle $\Delta^{+ + +}$ combined with the tree level 3-point MHV $(- - +)$ vertex from the classical action $S_{\mathrm{MHV}}[B]$ in Eq.~\eqref{eq:PartitionMHV}. The remaining contributions, i.e. those involving the one-loop triangle $\Delta^{+ + -}$ and the sword fish type terms, result in the Lorentz invariance violating contributions,
which is again an artifact of the regularization and must therefore be canceled by a counterterm. The counterterm has the following form \cite{CQT1}: 
\begin{center}
    \includegraphics[width=8cm]{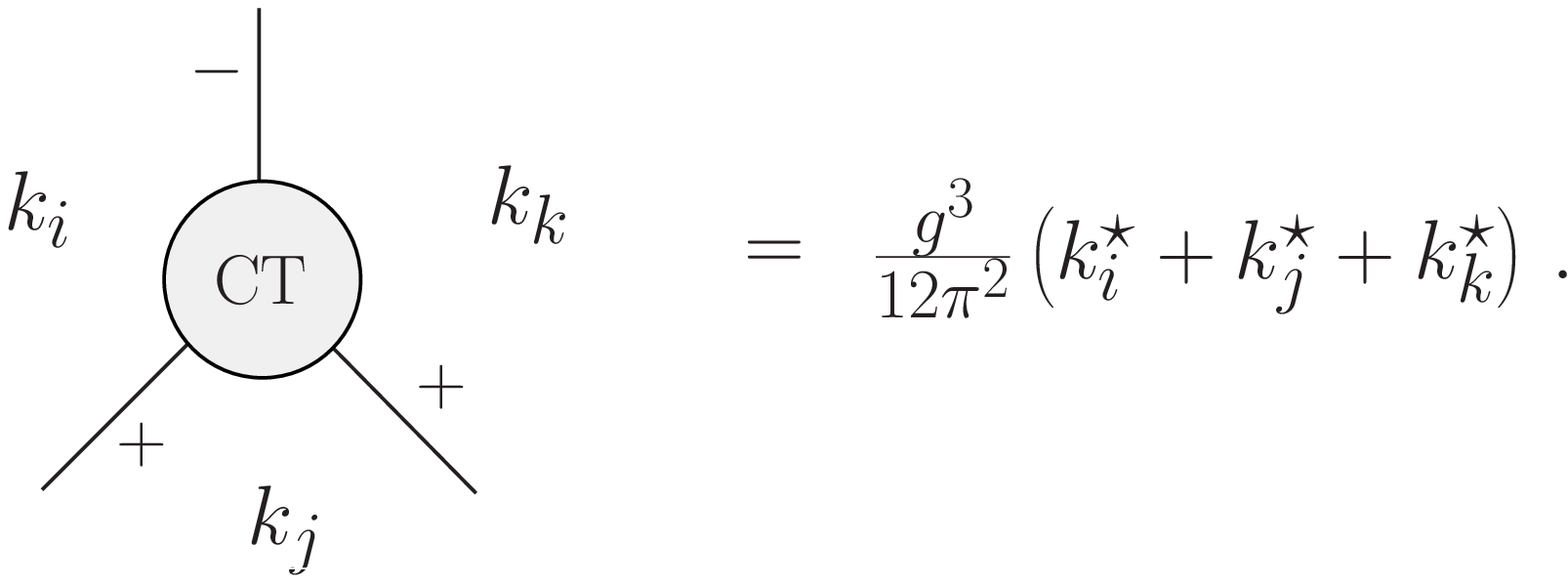}
\end{center}
 Substitution of the inverse Wilson lines Eq.~\eqref{eq:A_bull_exp}-\eqref{eq:A_star_exp} recovers  the result of 
\cite{CQT1}, which reads:
\begin{multline}
\mathcal{A}_{\mathrm{TS}}^{+ + + -}=\frac{-g^{4}p_{4}^{+}}{4 \pi^{2} p_{1}^{+} p_{2}^{+} p_{3}^{+}}\Bigg[\frac{{\tilde v}_{34}p_{4}^{+}p_{2}^{+}}{{\tilde v}_{21}^*}\Bigg\{\frac{22}{3} \ln \left(p_{12}^{2} e^{\gamma} \delta\right)-\frac{140}{9}-S_{3}^{q^{+}}\left(p_{1}, p_{2}\right)-S_{3}^{q^{+}}\left(-p_{4},-p_{3}\right)\\
+\frac{p_{1}^{+} p_{2}^{+}}{3 p_{12}^{+2}}\Bigg\}
+\frac{{\tilde v}_{41}p_{3}^{+}p_{1}^{+}}{{\tilde v}_{32}^*}\Bigg\{\frac{22}{3} \ln \left(p_{14}^{2} e^{\gamma} \delta\right)-\frac{140}{9}-S_{2}^{q^{+}}\left(-p_{4},-p_{23}\right)-S_{1}^{q^{+}+p_{4}^{+}}\left(p_{14}, p_{2}\right)+\frac{p_{2}^{+} p_{3}^{+}}{3 p_{14}^{+2}}\Bigg\}\Bigg] \\
+\frac{g^{4}p_{3}^{+}p_{2}^{+2}}{3 \pi^{2} p_{1}^{+}  p_{12}^{+2}} \frac{{\tilde v}_{12}^3 {\tilde v}_{34}^*}{p_{12}^{4}}+\frac{g^{4}p_{1}^{+ 2}p_{3}^{+ 2}}{3 \pi^{2} p_{2}^{+} p_{4}^{+} p_{14}^{+2}} \frac{{\tilde v}_{23}^3 {\tilde v}_{41}^*}{p_{14}^{4}}\,,
\label{eq:TS_contri_OL}
\end{multline}
where $S_{l}^{q^{+}}\!\left(p_{i}, p_{j}\right)$ is an infrared sensitive term whose explicit form is given in Appendix~\ref{sec:AppE}. Above, the subscript $\mathrm{TS}$ stands for "Triangle+Swordfish".

Finally, we have the box and the quartic diagrams. The quartic terms are the ones that involve the four point vertex and two triple gluon vertices in the loop (the last term in line 4 and first term in the line 5 in  Fiq.~\ref{fig:3plusminus}).
The box diagrams,  can be decomposed again in terms of triangle like diagrams, similar to the case discussed in the previous section for reducing $\square^{+ + + +}$. 
The box and quartic contributions do not get  contributions from the inverse Wilson lines, thus we simply use the result of \cite{CQT1}:
\begin{multline}
    \mathcal{A}_{\mathrm{BQ}}^{ + + + -}=\frac{ g^{4} }{2 \pi^{2}}\Bigg[ {-\left\{\frac{p_{1}^{+}p^+_{2\overline{3}}-3 p_{3}^{+}p^+_{23}}{6 p_{12}^{+} p_{23}^{+} p_{4}^{+} {\tilde v}_{14}{\tilde v}_{41}^*}+\frac{p_{3}^{+}(-3 p_{1}^{+} p_{12}^{+}+p_{3}^{+}p^+_{\overline{1}2})}{6 p_{1}^{+} p_{12}^{+2} p_{4}^{+} {\tilde v}_{12}{\tilde v}_{21}^*}\right\} {\tilde v}_{12}^2p_{2}^{+} } \\
+\left\{\frac{p_{1}^{+}p^+_{\overline{2}3}+p_{3}^{+} p_{23}^{+}}{6 p_{1}^{+} p_{3}^{+} p_{23}^{+2} {\tilde v}_{12}{\tilde v}_{21}^*}-\frac{p_{1}^{+}p^+_{\overline{2}3}+3 p_{3}^{+} p_{23}^{+}}{6 p_{12}^{+} p_{3}^{+} p_{23}^{+} p_{4}^{+} {\tilde v}_{14}{\tilde v}_{41}^*}\right\} {\tilde v}_{12}{\tilde v}_{34}p_2^+p_4^+\\
+\frac{p_{4}^{+}}{2p_{1}^{+} p_{2}^{+} p_{3}^{+}} \frac{{\tilde v}_{34}p_{4}^{+}p_{2}^{+}}{{\tilde v}_{21}^*}\Bigg\{\frac{11}{3} \ln \left(\delta e^{\gamma} p_{12}^{2}\right)-\frac{11}{3} \ln \left(\delta e^{\gamma} p_{14}^{2}\right)\\
     -S_{3}^{q^{+}}\left(p_{1}, p_{2}\right)-S_{3}^{q^{+}}\left(-p_{4},-p_{3}\right)+S_{2}^{q^{+}}\left(-p_{4},-p_{23}\right) +S_{1}^{q^{+}+p_{4}^{+}}\left(p_{14}, p_{2}\right)\\
-2 \sum_{q^{+}}\left[\frac{1}{q^{+}}+\frac{1}{p_{12}^{+}-q^{+}}\right] \ln \Bigg(\frac{q^{+}\left(p_{12}^{+}-q^{+}\right)}{p_{12}^{+ 2}} p_{12}^{2} e^{\gamma} \delta\Bigg)\\
+2 \sum_{q^{+}+p_4^+}\left[\frac{1}{q^{+}+p_4^+}+\frac{1}{p_1^{+}-q^{+}}\right] \ln \Bigg(\frac{(q^{+}+p_4^+)\left(p_1^{+}-q^{+}\right)}{p_{14}^{+2}} p_{14}^{2} e^{\gamma} \delta\Bigg)\Bigg\}\Bigg]\,,
\end{multline}
where we used the notation $p^+_{ij}=p_{i}^{+}+p_{j}^{+}$ and $p^+_{i\overline{j}}=p_{i}^{+}-p_{j}^{+}$.
Above, the subscript $\mathrm{BQ}$ stands for $\mathrm{"Box+Quartic"}$. 

The one-loop amplitude is give by the sum of all contributions:
\begin{equation}
    \mathcal{A}_{\mathrm{one-loop}}^{+ + + -} = \mathcal{A}_{\mathrm{SE}}^{+ + + -} + \mathcal{A}_{\mathrm{TS}}^{+ + + -} + \mathcal{A}_{\mathrm{BQ}}^{+ + + -}\,.
\end{equation}
Upon substitution of the above expressions, the divergent pieces explicitly cancel out and the final simplified expression after a bit of algebra reads
\begin{equation}
    \mathcal{A}_{\mathrm{one-loop}}^{+ + + -} = \frac{-g^{4}}{24 \pi^{2}} \frac{ p_3^{+}{\tilde v}_{13}^2}{p_1^+{\tilde v}_{14} {\tilde v}_{43}{\tilde v}^{\star}_{21} {\tilde v}^{\star}_{32}} (p_{12}^2 + p_{14}^2 )\,.
    \label{eq:3plus-oneloop}
\end{equation}
The above result is consistent with \cite{Bern:1991aq,Kunszt_1994} (modulo normalisation).

\section{Summary}
\label{sec:conclusion}

In this work we used  the approach of the one loop effective action to compute \emph{all} the  quantum corrections to the MHV theory.
In order to account for terms necessary to obtain all one-loop amplitudes, the one-loop effective action cannot be derived from the ordinary MHV action.
Instead, it has to be derived by applying the Mansfield's transformation \eqref{eq:MansfieldTransf2} to the one-loop effective action for the Yang-Mills theory \eqref{eq:Eaction}. 
This way, the self-dual interactions, instead of being completely removed, are "trapped" inside the functional determinant and provide the necessary non-MHV contributions after the field transformation is applied.
The side effect is that the MHV vertices appearing in a loop are not manifest; they however appear quite naturally after properly combining the field transformed loop contributions and the loop-tree connections.

As discussed in our previous works \cite{Kotko2017,Kakkad2020} the Mansfield's transformation of fields can be simply expressed in terms of the straight infinite Wilson line integrated over the slope, see \eqref{eq:WilsonLineBbullet}. In momentum space the Wilson line (or, more precisely, it's inverse) simply resums tree-level self-dual interactions. Therefore, the presented procedure reduces formally the number of diagrams, when compared with the Yang-Mills theory, because the inverse Wilson line solution \eqref{eq:A_bull_exp}-\eqref{eq:A_star_exp} takes care of tree level connections resulting from $(+ + -)$ triple gluon vertex.

Using the four-dimensional world-sheet regularization scheme we have explicitly checked that the one loop effective action is sufficient to compute $(+ + + +)$ and $(+ + + -)$ one-loop amplitudes, which could not be computed in the MHV theory. 

The one-loop effective action approach is well suited to develop quantum corrections in approaches based on classical field transformations, as the loop corrections are clearly separated from the classical part.
In particular, recently we developed an action  with no triple gluon vertices \cite{Kakkad:2021uhv}, which is based on canonically transformed classical MHV action, using Wilson line functionals. 
Similarly to the MHV theory this action has missing loop contributions, nevertheless we expect that applying the field transformations to the one-loop effective MHV action \eqref{eq:PartitionMHV} one should be able to  generate all the missing loop contributions in that theory. This, however, is planned for future work.

\section{Acknowledgments}
\label{sec:ack}
We thank Jacob Bourjaily and Radu Roiban for discussions. H.K. is supported by the National Science Centre, Poland grant no. 2021/41/N/ST2/02956. P.K. is supported by the National Science Centre, Poland grant no. 2018/31/D/ST2/02731.  A.M.S. is supported  by the U.S. Department of Energy Grant 
 DE-SC-0002145 and  in part by  National Science Centre in Poland, grant 2019/33/B/ST2/02588.

\bibliographystyle{JHEP}
\bibliography{library}

\newpage
\appendix

\section{Computing amplitudes from Effective Action}
\label{sec:AppA}

In the following Appendix we review some basic facts about effective action. In particular, we remind how scattering amplitudes are computed using an effective action. 
This short review is similar to standard textbook expositions, see for example \cite{Peskin:1995ev}.

Let us start with a generic definition of an effective action through a Legendre transformation:
\begin{equation}
    \Gamma[A_c]=W[J]-J_IA_c^{I} \,,
    \label{eq:gam_gen}
\end{equation}
where we use the abstract index notation.
Then, it follows that
\begin{equation}
     \frac{\delta \Gamma[A_c]}
    {\delta A_c^{I}} = -J_I \, , \quad \frac{\delta W[J]}
    {\delta J_I} = A_c^{I} \, .
    \label{eq:EOM_EA}
\end{equation}

The most basic object is the two point connected Green's function, which is a second derivative of $W$ with respect to the sources. In order to relate it to the effective action, we use the following relation:
\begin{equation}
    \frac{\delta}
    {\delta A_c^{J}}\left(\frac{\delta W[J]}
    {\delta J_I} \right)=\frac{\delta J_K}
    {\delta A_c^{J}}\left(\frac{\delta^2 W[J]}
    {\delta J_K \delta J_I} \right) = - \frac{\delta^2 \Gamma[A_c]}
    {\delta A_c^{J} \delta A_c^{K}}\left(\frac{\delta^2 W[J]}
    {\delta J_K \delta J_I} \right) = \delta^{I}_J \, .
    \label{eq:EOM_2}
\end{equation}
\begin{figure}
    \centering
    \includegraphics[width=7cm]{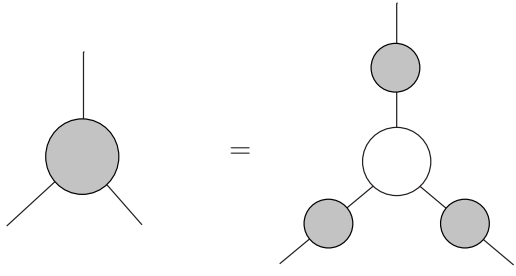}
    \caption{
    Each dark blob in this figure represents the sum of connected diagrams, whereas the white blob on the right represents the third derivative of $\Gamma[A_c]$ \eqref{eq:gam_gen}. We see that the third derivative of $\Gamma[A_c]$ is just the connected 
correlation function with all three full propagators removed, i.e. the one-particle-irreducible three-point function.}
    \label{fig:3_point_FR}
\end{figure}
It follows then, that the second order derivative of the effective action is the inverse of the two point connected Green's function. More precisely
\begin{equation}
    \frac{\delta^2 W[J]}
    {\delta J_K \delta J_I} = - \left(\frac{\delta^2 \Gamma[A_c]}
    {\delta A_c^{I} \delta A_c^{K}} \right)^{-1} \, .
    \label{eq:w2_g2}
\end{equation}
Thus,  the second order derivative of the effective action is sometimes called  an inverse propagator. 
When tadpoles are nonzero, the relation above has to be modified either by explicitly including tadpoles or by modifying the definition of a proper 2-point 1PI function.

Differentiating Eq. \eqref{eq:w2_g2} one more time, we get
\begin{multline}
    \frac{\delta^3 W[J]}
    {\delta J_L \delta J_K \delta J_I} = - \frac{\delta A_c^{M}}
    {\delta J_L} \frac{\delta}
    {\delta A_c^{M}} \left[ \left(\frac{\delta^2 \Gamma[A_c]}
    {\delta A_c^{I} \delta A_c^{K}} \right)^{-1} \right] \\
    =- \left(\frac{\delta^2 \Gamma[A_c]}
    {\delta A_c^{M} \delta A_c^{L}} \right)^{-1}  \left(\frac{\delta^2 \Gamma[A_c]}
    {\delta A_c^{N} \delta A_c^{K}} \right)^{-1} 
    \left(\frac{\delta^2 \Gamma[A_c]} 
    {\delta A_c^{P} \delta A_c^{I}} \right)^{-1} 
    \frac{\delta^3 \Gamma[A_c]}
    {\delta A_c^{M} \delta A_c^{N} \delta A_c^{P}} \, .
    \label{eq:w3_g3}
\end{multline}
Thus, the triple functional derivative of the effective action $\Gamma[A_c]$ gives the amputated 3-point connected Green's function, i.e. two point connected Green's function are removed from each leg (see Fig~\ref{fig:3_point_FR}).

\begin{figure}
    \centering
    \includegraphics[width=13.5cm]{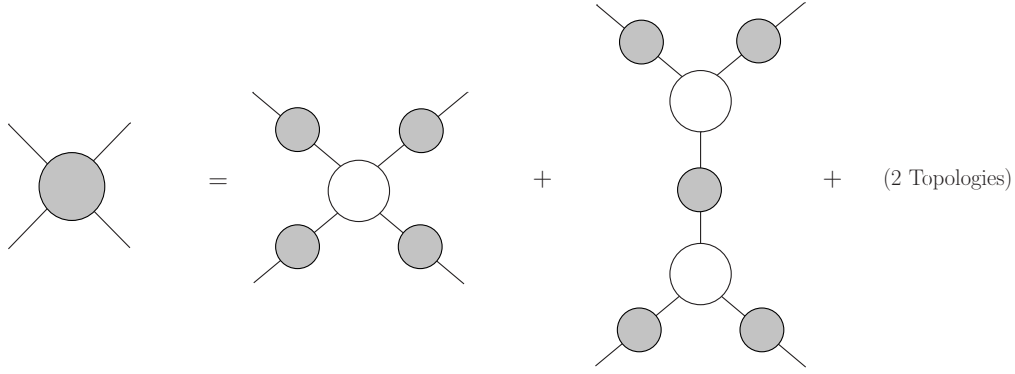}
    \caption{
    Each dark blob in this figure represents the sum of connected diagrams, whereas the white blobs on the right represents the fourth derivative of $\Gamma[A_c]$ in the first term and the third derivative of $\Gamma[A_c]$ in the second. There are two more terms just like the second term in the right. "2 Topologies" represent these terms.}
    \label{fig:4_point_FR}
\end{figure}

In the similar fashion we can derive: 
\begin{multline}
    \frac{\delta^4 W[J]}
    {\delta J_Q \delta J_L \delta J_K \delta J_I} = \frac{\delta A_c^{R}}
    {\delta J_Q} \frac{\delta}
    {\delta A_c^{N}} \left( \frac{\delta^3 W[J]}
    {\delta J_L \delta J_K \delta J_I} \right) \\
    =-\left(\frac{\delta^2 \Gamma[A_c]}
    {\delta A_c^{R} \delta A_c^{Q}} \right)^{-1} \left(\frac{\delta^2 \Gamma[A_c]}
    {\delta A_c^{M} \delta A_c^{L}} \right)^{-1} \frac{\delta^3 \Gamma[A_c]}
    {\delta A_c^{R} \delta A_c^{M} \delta A_c^{T}} \left(\frac{\delta^2 \Gamma[A_c]}     {\delta A_c^{S} \delta A_c^{T}} \right)^{-1}  \left(\frac{\delta^2 \Gamma[A_c]}
    {\delta A_c^{N} \delta A_c^{K}} \right)^{-1} \\ \times \frac{\delta^3 \Gamma[A_c]}
    {\delta A_c^{S} \delta A_c^{N} \delta A_c^{P}} \left(\frac{\delta^2 \Gamma[A_c]}     {\delta A_c^{P} \delta A_c^{I}} \right)^{-1}
   + \mathrm{2 \,\, Topologies\,\,}\\ - \left(\frac{\delta^2 \Gamma[A_c]}
    {\delta A_c^{R} \delta A_c^{Q}} \right)^{-1} \left(\frac{\delta^2 \Gamma[A_c]}
    {\delta A_c^{M} \delta A_c^{L}} \right)^{-1} \frac{\delta^4 \Gamma[A_c]}
    {\delta A_c^{R} \delta A_c^{M} \delta A_c^{N}\delta A_c^{P}} \left(\frac{\delta^2 \Gamma[A_c]}
    {\delta A_c^{N} \delta A_c^{K}} \right)^{-1}  \left(\frac{\delta^2 \Gamma[A_c]} 
    {\delta A_c^{P} \delta A_c^{I}} \right)^{-1}
    \label{eq:w4_g4}
\end{multline}
This is diagrammatically represented in Fig.~\ref{fig:4_point_FR}. 
Relations for higher derivatives of  $\Gamma[A_c]$ can be computed in a similar fashion. In summary, the derivatives of the effective action \eqref{eq:gam_gen} give the \textit{one-particle irreducible } (1PI) contributions. 

\begin{figure}
    \centering
    \includegraphics[width=14cm]{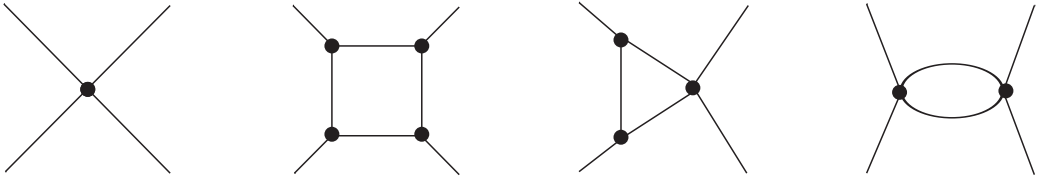}
    \caption{The figure represents the geometries obtained via the fourth derivative of $\Gamma[A_c]$. The first term is just the four point vertex. The remaining three are four point contributions obtained from the log term. For simplicity we suppress the symmetry factors, sign etc.}
    \label{fig:gamma_4V}
\end{figure}

\begin{figure}
    \centering
    \includegraphics[width=14cm]{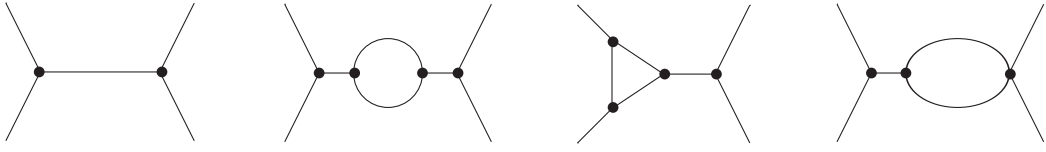}
    \caption{The figure represents the geometries obtained via combining the third derivative of $\Gamma[A_c]$ via a propagator or bubble diagram. For simplicity we suppress the symmetry factors, sign etc.}
    \label{fig:gamma_3VC}
\end{figure}

For example, consider 4-point amplitude up to one-loop, in a theory consisting of 3 and 4 point vertices.  There are two types of contributions. First, consider 1PI contributions obtained  through the fourth derivative of $\Gamma[A_c]$. These are shown in Fig.~\ref{fig:gamma_4V}
The second type of contribution arises by combining the third derivative of $\Gamma[A_c]$ via a propagator or bubble diagram. These are shown in Fig.~\ref{fig:gamma_3VC}.
These diagrams exhaust the set of possible geometries contributing to a four point amplitude up to one-loop.

Let us summarize the use of effective action to compute amplitudes:
\begin{itemize}
   \item The functional derivative of the effective action represent the vertices.
   \begin{equation}
       \Bigg[\frac{\delta^n \Gamma[A_c]}
    {\delta A_c^{I} \delta A_c^{J} \dots \delta A_c^{K}}\Bigg]_{A_c= 0} \,.
   \end{equation}
   \item The vertices can be connected together using the propagator as well as the bubbles i.e two point connected Green's function.
\end{itemize}
\section{Four point MHV}
\label{sec:AppB}

In this appendix we show the details that lead to the following result:
\begin{center}
    \includegraphics[width=13cm]{mhv_4pr.eps}
\end{center}
As pointed out before, we keep the loop terms un-traced i.e. we cut open the loop. The double line represents a propagator.

For simplicity, we name the legs $(- - + +)$ as 1, 2, 3  and 4  anti-clockwise. We associate  the color and momentum $\left\{b_i, {p}_{i}\right\}$ respectively to each leg, where all the external momenta are considered outgoing. Let us focus on the color ordered case.

The first term is the un-traced 4 point Yang-Mills vertex which, using Eq.~\eqref{eq:YM_LC_action}, in momentum space reads:
\begin{equation}
    D_1 = -\frac{1}{2}\frac{4}{p_1^2} g^2 \frac{p_1^+ p_3^+ + p_2^+ p_4^+}{(p_1^+ + p_4^+)^2} \,\, \mathrm{Tr}(t^{b_1}t^{b_2}t^{b_3}t^{b_4}) \,.
\end{equation}
The factor of 4 arises due to the second order functional derivative of this vertex. The second term is a tree level connection of the un-traced $(+ + -)$ tadpole from the logarithm term in Eq.~\eqref{eq:PartitionMHV} with the 3 point MHV $(+ - -)$ vertex from the classical action $S_{MHV}[B]$, following the rules discussed in Appendix \ref{sec:AppA}. This is given as:
\begin{equation}
    D_2 = -\frac{1}{p_1^2}g^2 2\Bigg(\frac{p^{\star}}{p^+} - \frac{p_4^{\star}}{p_4^+}\Bigg)p_1^+ \frac{1}{p^2} \,2\, \Bigg(\frac{p^{\bullet}}{p^+} - \frac{p_2^{\bullet}}{p_2^+}\Bigg)p_3^+\,\, \mathrm{Tr}(t^{b_1}t^{b_2}t^{b_3}t^{b_4}) \,,
\end{equation}
Using $p=p_1+p_4=-(p_2+p_3)$, and $p^2=2(p^+p^- - p^{\bullet}p^{\star})$, we get:
\begin{equation}
    D_2 = \frac{4}{p_1^2}g^2 \frac{1}{2} \frac{(p_1^+)^2 p_3^+ {\tilde v}^{\star}_{2(23)}}{p_{14}^+ p_4^+ p_2^+ {\tilde v}^{\star}_{14}}\,\, \mathrm{Tr}(t^{b_1}t^{b_2}t^{b_3}t^{b_4}) \,.
\end{equation}
The third term involves substitution of the second order expansion of the $\hat{A}_c^{\star}[B(x)]$ field to the  un-traced $(+ - -)$ tadpole from the logarithm term in Eq.~\ref{eq:PartitionMHV}, resulting in the kernel ${\widetilde \Omega}_{2}^{a b_2 \left \{b_3 \right \} }(\mathbf{p}; \mathbf{p}_2 ,\left \{ \mathbf{p}_3 \right \})$. With a bit of algebra, we get:
\begin{equation}
    D_3 = \frac{4}{p_1^2}g^2 \frac{1}{2} \frac{p_2^+ p_4^+ {\tilde v}^{\star}_{(14)1}}{p_{14}^+ p_{23}^+ {\tilde v}^{\star}_{32}}\,\, \mathrm{Tr}(t^{b_1}t^{b_2}t^{b_3}t^{b_4}) \,.
\end{equation}
The final is the un-traced $(- +)$ bubble term from the logarithm term in Eq.~\eqref{eq:PartitionMHV}. This term reads:
\begin{equation}
    D_4 = -\frac{4}{p_1^2}g^2 \frac{1}{2} \frac{p_{12}^+ p_{34}^+ {\tilde v}^{\star}_{21}}{p_2^+ p_3^+ {\tilde v}^{\star}_{43}}\,\, \mathrm{Tr}(t^{b_1}t^{b_2}t^{b_3}t^{b_4}) \,.
\end{equation}
The sum of all these contributions, with a bit of algebra, reads:
\begin{equation}
    D_1 +  D_2 +  D_3 + D_4 = \frac{4}{p_1^2}\frac{g^2}{2}  \left(\frac{p_{1}^{+}}{p_{2}^{+}}\right)^{2}
\frac{\widetilde{v}_{21}^{*4}}{\widetilde{v}_{14}^{*}\widetilde{v}_{43}^{*}\widetilde{v}_{32}^{*}\widetilde{v}_{21}^{*}}\,\, \mathrm{Tr}(t^{b_1}t^{b_2}t^{b_3}t^{b_4}) \,.
\end{equation}
The right hand side of the expression above is an un-traced 4-point MHV vertex with a propagator attached to a leg. 

\section{Gluon Self-Energy $\Pi^{+ +}$ and Triangle contribution in CQT regularisation}
\label{sec:App_CQT_calculations}

\begin{figure}
    \centering
 \includegraphics[width=3.5cm]{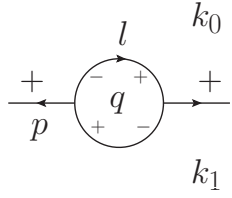}
    \caption{\small 
    The gluon $(+ +)$ self-energy diagram. $q$ is the region momentum associated with the loop and $k_i$ with the external region. $l$ is the loop momentum and $p$ is the external line momentum. These are related to the region momenta via: $l=q-k_0$ and $p= k_0-k_1$ respectively.}
    \label{fig:2_ct}
\end{figure}

In this Appendix, we review how basic loop calculations are carried out in the CQT scheme. 

Let us first discuss the details of the $(++)$ gluon self-energy calculation. The contributing diagram is shown in Fig.~\ref{fig:2_ct} where, $p$ and $l$ represent external line and loop momentum respectively.
In terms of the line momenta, the expression for $\Pi^{+ +}$ reads:
\begin{align}
\Pi^{++} &=8 g^{2} \int \frac{\mathrm{d}^{4} l}{(2 \pi)^{4}}\Bigg(\frac{p^{\star}}{p^+} - \frac{l^{\star}}{l^+}\Bigg)(p+l)^+ \frac{1}{l^2} \,\frac{1}{{(p+l)}^2}\, \Bigg(\frac{p^{\star}}{p^+} - \frac{{(p+l)}^{\star}}{{(p+l)}^+}\Bigg)l^+ \,, \\
&=\frac{g^{2}}{2 \pi^{4}} \int \mathrm{d}^{4} l \frac{1}{\left(p^{+}\right)^{2}}\left(p^{+} l^{\star}-l^{+} p^{\star}\right)\left(p^{+}\left(p^{\star}+l^{\star}\right)-\left(p^{+}+l^{+}\right) p^{\star}\right) \frac{1}{l^{2}(p+l)^{2}} ,
\end{align}
where we suppress the color factors. 
Note that the $1/l^+$ singularity gets canceled and thus we can keep the plus component continuous.

Using \eqref{eq:LM_con}, the line momenta can be re-written in terms of the region momenta $k_0$, $k_1$ and $q$ as:
\begin{equation}
    p= k_0-k_1 \,,\quad \quad l=q-k_0 \,.
    \label{eq:line_def2p}
\end{equation}
Introducing the regulator and using the Schwinger representation for the propagators
\begin{equation}
    \frac{e^{-\delta \boldsymbol{q}^{2}}}{\left(q-k_{0}\right)^{2}\left(q-k_{1}\right)^{2}}=\int_0^{\infty} d t_{1} d t_{2} e^{-t_{1}\left(q-k_{0}\right)^{2}-t_{2}\left(q-k_{1}\right)^{2}-\delta \mathbf{q}^{2}} \,,
\end{equation}
we have:
\begin{align}
\Pi^{++}=& \frac{g^{2}}{2 \pi^{4}} \int_{0}^{\infty} \mathrm{d} t_{1} \mathrm{~d} t_{2} \int \mathrm{d}^{4} q \frac{1}{\left(k_{0}^{+}\right)^{2}} e^{-t_{1}(q-k_0)^{2}-t_{2}\left(q-k_1\right)^{2}-\delta \mathbf{q}^{2}}  \\
& \times\left[k_{0}^{+}\left(q^{\star}-k_0^{\star}\right)-\left(q^{+}-k_{0}^{+}\right)\left(k_{0}^{\star}-k_{1}^{\star}\right)\right]\left[k_{0}^{+}\left(q^{\star}-k_1^{\star}\right)-q^{+}\left(k_{0}^{\star}-k_1^{\star}\right)\right] \,,
\end{align}
where we set $k_{1}^{+}=0$ owing to the translational invariance along the $+$ component of the region momentum. 
The $q^{-}$ integration results in $\pi \delta\left(\left(t_{1}+t_{2}\right) q^{+}-t_{2} p^{+}\right)$. This can be used to integrate out $q^{+}$. We are left with the integral over transverse momentum of the Gaussian type. Integrating out the loop region momenta  and  performing the change of variable $T=t_{1}+t_{2}, \alpha=t_{1} /\left(t_{1}+t_{2}\right)$, we get
\begin{equation}
   \Pi^{++}=\frac{g^{2}}{2 \pi^{2}} \int_{0}^{1} \mathrm{~d} \alpha \int_{0}^{\infty} \mathrm{d} T \frac{ \delta^{2}\,\left[\alpha k_1^{\star}+(1-\alpha) k_0^{\star}\right]^{2}}{(T+\delta)^{3}} e^{-T \alpha(1-\alpha) p^{2}-\frac{\delta T}{T+\delta}\left(\alpha \mathbf{k}_0+(1-\alpha) \mathbf{k}_1\right)^{2}}  \,.
\end{equation}
Integrating out $T$, in the limit $\delta \longrightarrow 0$ we get:
\begin{equation}
    \Pi^{+ +}= \frac{g^{2}}{4 \pi^{2}} \int_{0}^{1} \mathrm{~d} \alpha \left[\alpha k_1^{\star}+(1-\alpha) k_0^{\star}\right]^{2}\,.
\end{equation}
Finally, the integration over $\alpha$ gives:
\begin{equation}
    \Pi^{+ +}= \frac{g^{2}}{12 \pi^{2}} \left[k_{0}^{\star 2}+k_{1}^{\star 2}+k_{0}^{\star} k_{1}^{\star}\right]\,.
    \label{eq:++GSE_ap}
\end{equation}

 \begin{figure}
    \centering
 \includegraphics[width=4cm]{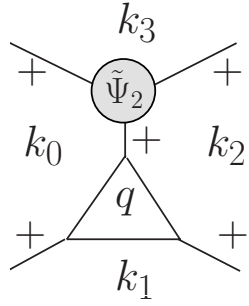}
    \caption{\small 
The $\Delta^{+ + + +}_{12}$ triangular one-loop diagram. $q$ is the region momentum associated with the loop and $k_i$ with the external region. ${\tilde \Psi}_2$ represents the inverse Wilson line kernel Eq.~\eqref{eq:psi_kernel} expanded to second order. }
    \label{fig:4_tri}
\end{figure}

 \begin{figure}
    \centering
 \includegraphics[width=4cm]{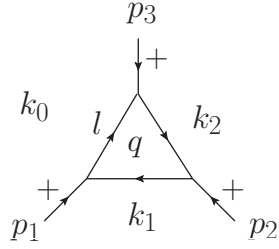}
    \caption{\small 
    The $(+ + +)$ triangle one-loop diagram. $q$ is the region momentum associated with the loop and $k_i$ with the external region. $l$ is the loop momentum and $p_i$ are the external line momenta. These are related to the region momenta via: $l=q-k_0$ and $p_1= k_1-k_0$, $p_2= k_2-k_1$, $p_3= k_0-k_2$ respectively.}
    \label{fig:3_tri}
\end{figure}

Next, let us consider the triangular contribution to the four point $(+ + + +)$ amplitude shown in Fig.~\ref{fig:4_tri}. We first start with just the triangle $\Delta^{+ + +}$ sub-diagram shown in Fig.~\ref{fig:3_tri}, where the arrows represent the orientation of the line momenta. 
In terms of the line momenta, the expression for $\Delta^{+ + +}$ reads:
\begin{multline}
\Delta^{+ + +} = -g^{3}16 \int \frac{\mathrm{d}^{4} l}{(2 \pi)^{4}}\Bigg(\frac{p_1^{\star}}{p_1^+} - \frac{l^{\star}}{l^+}\Bigg)(-p_1+l)^+ \frac{1}{l^2}\,\frac{1}{{(-p_1+l)}^2} \,\frac{1}{{(p_3+l)}^2}\,\\ \Bigg(\frac{p_3^{\star}}{p_3^+} - \frac{{(p_3+l)}^{\star}}{{(p_3+l)}^+}\Bigg)l^+ \Bigg(\frac{p_2^{\star}}{p_2^+} - \frac{{(-p_1+l)}^{\star}}{{(-p_1+l)}^+}\Bigg)(l+p_3)^+ \,,
\end{multline}
where we suppress the color factors as before. The above expression simplifies to:
\begin{multline}
  \Delta^{+ + +}  =-\frac{g^{3}}{\pi^{4}} \int \mathrm{d}^{4} l \frac{1}{p_1^{+}p_2^{+}p_3^{+}}\left(p_1^{*} l^{+}-l^{*} p_1^{+}\right)\left(p_2^{*}\left(-p_1^{+}+l^{+}\right)-\left(-p_1^{*}+l^{*}\right) p_2^{+}\right) \\
 \left(p_3^{*}\left(p_3^{+}+l^{+}\right)-\left(p_3^{*}+l^{*}\right) p_3^{+}\right) \frac{1}{l^{2}{(-p_1+l)}^2(p_3+l)^{2}} .
\end{multline}
Since the $1/l^+$ singularity cancels out, we can again keep the plus component continuous.
Following \eqref{eq:LM_con}, the line momenta are expressed in terms of the region momenta as:
\begin{equation}
    p_1= k_1-k_0 \,,\quad p_2= k_2-k_1\,, \quad p_3= k_0-k_2\,, \quad l=q-k_0 \,.
    \label{eq:line_def}
\end{equation}
Expressing the diagram in terms of the region momenta we get:
\begin{multline}
  \Delta^{+ + +}  =-\frac{g^{3}}{\pi^{4}} \int \mathrm{d}^{4} q \frac{1}{p_1^{+}p_2^{+}p_3^{+}}\left((k_1-k_0)^{*} \left(q-k_0\right)^{+}-(q-k_0)^{*} (k_1-k_0)^{+}\right)\\
  \left((k_2-k_1)^{*}\left(q-k_1\right)^+-\left(q-k_1\right)^* (k_2-k_1)^{+}\right) \\
 \left((k_0-k_2)^{*}\left(q-k_2\right)^+-\left(q-k_2\right)^* (k_0-k_2)^{+}\right)\\ \int_0^{\infty} d t_{1} d t_{2} d t_{3} e^{-t_{1}\left(q-k_{0}\right)^{2}-t_{2}\left(q-k_{1}\right)^{2}-t_{3}\left(q-k_{2}\right)^{2}-\delta \mathbf{q}^{2}}\, .
\end{multline}
As previously shown, integrating out the loop momenta involves three steps: first integrate out $q^-$ to get a delta function in $q^+$, then use it to integrate out the $q^+$. Finally, complete the square in the exponential and perform the Gaussian integral in $\mathbf{q}$. Following this we get:
\begin{align}
\Delta^{+ + +}=&-\frac{g^{3}}{\pi^{2}} \int_{0}^{\infty} \frac{d t_{1} d t_{2} d t_{3}}{p_1^{+}p_2^{+}p_3^{+}t_{123}\left(t_{123}+\delta\right)} \exp \left(-\frac{\delta\left(\boldsymbol{k}_{0} t_{3}+\boldsymbol{k}_{1} t_{1}+\boldsymbol{k}_{2} t_{2}\right)^{2}}{t_{123}\left(t_{123}+\delta\right)}\right) \nonumber\\
& \exp \left(-\frac{t_{1} t_{3}\left(k_{1}-k_{0}\right)^{2}+t_{1} t_{2}\left(k_{1}-k_{2}\right)^{2}+t_{2} t_{3}\left(k_{2}-k_{0}\right)^{2}}{t_{123}}\right) \nonumber\\
& \left(\frac{t_{1} {\tilde v}_{12}p_2^+}{ t_{123}}-\frac{\delta\left(t_{1} k_{1}^{\star}+t_{2} k_{2}^{\star}+t_{3} k_{0}^{\star}\right)}{t_{123}\left(\delta+t_{123}\right)}\right) \left(\frac{t_{2} {\tilde v}_{12}p_2^+}{ t_{123}}-\frac{\delta\left(t_{1} k_{1}^{\star}+t_{2} k_{2}^{\star}+t_{3} k_{0}^{\star}\right)}{t_{123}\left(\delta+t_{123}\right)}\right)\nonumber\\
&\left(\frac{t_{3} {\tilde v}_{12}p_2^+}{ t_{123}}-\frac{\delta\left(t_{1} k_{1}^{\star}+t_{2} k_{2}^{\star}+t_{3} k_{0}^{\star}\right)}{t_{123}\left(\delta+t_{123}\right)}\right)\,,
\end{align}
where $t_{123}=t_{1}+t_{2}+t_{3}$. 

In the $\delta \rightarrow 0$ limit we get:
\begin{multline}
\Delta^{+ + +}  =-\frac{g^{3}}{\pi^{2}} \int_{0}^{\infty} \frac{d t_{1} d t_{2} d t_{3}}{p_{1}^{+} p_{2}^{+} p_{3}^{+} t_{123}^{5}} t_{1} t_{2} t_{3}\left({\tilde v}_{12}p_2^+\right)^{3} \\
\exp \left(-\frac{t_{1} t_{3}\left(k_{1}-k_{0}\right)^{2}+t_{1} t_{2}\left(k_{1}-k_{2}\right)^{2}+t_{2} t_{3}\left(k_{2}-k_{0}\right)^{2}}{t_{123}}\right) \,.
\end{multline}
Performing the change of variables: 
$T=t_{1}+t_{2}+t_{3}$,  $\alpha=t_{1} /\left(t_{1}+t_{2}+t_{3}\right)$, $\beta=t_{2} /\left(t_{1}+t_{2}+t_{3}\right)$, and integrating out $T$, we get:
\begin{multline}
    \Delta^{+ + +}  =-\frac{g^{3}}{\pi^{2}} \frac{\left({\tilde v}_{12}p_2^+\right)^{3}}{p_{1}^{+} p_{2}^{+} p_{3}^{+}}\\ \int_{\alpha+\beta<1} d \alpha \,\,d \beta \frac{\alpha \beta(1-\alpha-\beta)}{\alpha(1-\alpha-\beta)\left(k_{1}-k_{0}\right)^{2}+\alpha \beta\left(k_{1}-k_{2}\right)^{2}+\beta(1-\alpha-\beta)\left(k_{2}-k_{0}\right)^{2}} \,.
\end{multline}
Since, we want to obtain the contribution of this triangle to on-shell scattering at one loop, we put the legs 1,2 on-shell: $p_{1}^{2}=0 $, $p_{2}^{2}=0$. With this we get:
\begin{equation}
    \Delta^{+ + +}  =-\frac{g^{3}}{6 \pi^{2}} \frac{\left({\tilde v}_{12}p_2^+\right)^{3}}{p_{1}^{+} p_{2}^{+} p_{3}^{+} p_3^{2}}\,.
\end{equation}

Now, let us use this to get the triangular contribution shown in Fig.~\ref{fig:4_tri}. For this we need to substitute the inverse Wilson line kernel ${\tilde \Psi}_2$ using Eq.~\eqref{eq:A_bull_exp}. Doing this we finally get
\begin{equation}
    \Delta^{+ + + +}_{12}  =-\frac{g^{4}}{12 \pi^{2}} \frac{\left({\tilde v}_{12}p_2^+\right)^{3}p_{3}^{+}}{p_{1}^{+} p_{2}^{+} p_{3}^{+}p_{4}^{+} p_{34}^{2}{\tilde v}^*_{34}}\,.
\end{equation}

\section{Expression for $S_{l}^{q^{+}}\left(p_{i}, p_{j}\right)$}
\label{sec:AppE}

In this Appendix, for completeness, we list expressions for $S_{l}^{q^{+}}\left(p_{i}, p_{j}\right)$ with $i=1$ and $j=2$, which are needed to obtain $(+\dots +-)$ one loop amplitude. These expressions were  derived in \cite{CQT1,CQTnotes} and they have the following form
\begin{multline}
S_{1}^{q^{+}}\left(p_{1}, p_{2}\right)=\sum_{q^{+}<p_{1}^{+}}\left\{\left[\frac{2}{q^{+}}+\frac{1}{p_{12}^{+}-q^{+}}+\frac{1}{p_{1}^{+}-q^{+}}\right]\left(\ln \left(\delta p_{1}^{2} e^{\gamma}\right)+\ln \frac{q^{+}}{p_{1}^{+}}\right)\right.\\
\left.+\left[\frac{2}{p_{1}^{+}-q^{+}}-\frac{1}{p_{12}^{+}-q^{+}}+\frac{1}{q^{+}}\right] \ln \frac{p_{1}^{+}-q^{+}}{p_{1}^{+}}\right\} \\
+\sum_{q^{+}>p_{1}^{+}}\left\{\left[\frac{1}{q^{+}}+\frac{2}{p_{12}^{+}-q^{+}}+\frac{1}{q^{+}-p_{1}^{+}}\right]\left(\ln \left(\delta p_{1}^{2} e^{\gamma}\right)+\ln \frac{p_{12}^{+}-q^{+}}{p_{2}^{+}}\right)\right. \\
+\sum_{q^{+} \neq p_{1}^{+}}\left[\frac{1}{q^{+}}+\frac{2}{p_{12}^{+}-q^{+}}+\frac{1}{q^{+}-p_{1}^{+}}\right] \ln \frac{p_{12}^{+}-q^{+}}{p_{12}^{+}}\,,
\end{multline}

\begin{multline}
    S_{2}^{q^{+}}\left(p_{1}, p_{2}\right)=\sum_{q^{+} \neq p_{1}^{+}}\left[\frac{2}{q^{+}}+\frac{1}{p_{12}^{+}-q^{+}}+\frac{1}{p_{1}^{+}-q^{+}}\right] \ln \frac{q^{+}}{p_{12}^{+}}\\
+\sum_{q^{+}<p_{1}^{+}}\left\{\left[\frac{2}{q^{+}}+\frac{1}{p_{12}^{+}-q^{+}}+\frac{1}{p_{1}^{+}-q^{+}}\right]\left(\ln \left(\delta p_{2}^{2} e^{\gamma}\right)+\ln \frac{q^{+}}{p_{1}^{+}}\right)\right\}\\
+\sum_{q^{+}>p_{1}^{+}}\left\{\left[\frac{1}{q^{+}}+\frac{2}{p_{12}^{+}-q^{+}}+\frac{1}{q^{+}-p_{1}^{+}}\right]\left(\ln \left(\delta p_{2}^{2} e^{\gamma}\right)+\ln \frac{p_{12}^{+}-q^{+}}{p_{2}^{+}}\right)\right.
\\
\left. +\left[\frac{2}{q^{+}-p_{1}^{+}}+\frac{1}{p_{12}^{+}-q^{+}}-\frac{1}{q^{+}}\right] \ln \frac{q^{+}-p_{1}^{+}}{p_{2}^{+}}\right\}\,,
\end{multline}

\begin{multline}
    S_{3}^{q^{+}}\left(p_{1}, p_{2}\right)=\sum_{q^{+}<p_{1}^{+}}\left\{\left[\frac{2}{q^{+}}+\frac{1}{p_{12}^{+}-q^{+}}+\frac{1}{p_{1}^{+}-q^{+}}\right]\left(\ln \left(\delta p_{12}^{2} e^{\gamma}\right)+\ln \frac{q^{+}}{p_{12}^{+}}\right)\right.\\
+\left[\frac{1}{q^{+}}+\frac{2}{p_{12}^{+}-q^{+}}+\frac{1}{q^{+}-p_{1}^{+}}\right] \ln \frac{p_{12}^{+}-q^{+}}{p_{12}^{+}}\\
\left.+\left[\frac{2}{p_{1}^{+}-q^{+}}-\frac{1}{p_{12}^{+}-q^{+}}+\frac{1}{q^{+}}\right] \ln \frac{p_{1}^{+}-q^{+}}{p_{1}^{+}}\right\}\\
+\sum_{q^{+}>p_{1}^{+}}\left\{\left[\frac{1}{q^{+}}+\frac{2}{p_{12}^{+}-q^{+}}+\frac{1}{q^{+}-p_{1}^{+}}\right]\left(\ln \left(\delta p_{12}^{2} e^{\gamma}\right)+\ln \frac{p_{12}^{+}-q^{+}}{p_{12}^{+}}\right)\right.\\
+\left[\frac{2}{q^{+}}+\frac{1}{p_{12}^{+}-q^{+}}+\frac{1}{p_{1}^{+}-q^{+}}\right] \ln \frac{q^{+}}{p_{12}^{+}}\\
\left.+\left[\frac{2}{q^{+}-p_{1}^{+}}+\frac{1}{p_{12}^{+}-q^{+}}-\frac{1}{q^{+}}\right] \ln \frac{q^{+}-p_{1}^{+}}{p_{2}^{+}}\right\}\,.
\end{multline}

\end{document}